\documentclass[twocolumn]{aastex62}

\usepackage{longtable}
\usepackage{rotating}
\usepackage{float}
\usepackage{lineno}
\graphicspath{{./}{figures/}}

\received{}
\revised{}
\accepted{}

\shorttitle{CH$_3$OD in Prestellar Cores}
\shortauthors{Kulterer et al.}

\begin{document}

\title{First Detection of CH$_3$OD in Prestellar Cores}

\author[0000-0002-7609-3966]{Beatrice M. Kulterer}
\affiliation{Center for Astrophysics | Harvard \& Smithsonian, Cambridge, MA 02138, USA}
\author{Asunción Fuente}
\affiliation{Centro de Astrobiología (CAB), CSIC-INTA, Ctra. de Torrejón a Ajalvir, km 4, 28850 Torrejón de Ardoz, Spain}
\author{Maria N. Drozdovskaya}
\affiliation{Physikalisch-Meteorologisches Observatorium Davos und Weltstrahlungszentrum (PMOD/WRC), Dorfstrasse 33, CH-7260, Davos Dorf, Switzerland}
\author{Silvia Spezzano}
\affiliation{Max-Planck-Institut für Extraterrestrische Physik, Giessenbachstrasse 1, 85748 Garching, Germany}
\author{Gisela Esplugues}
\affiliation{Observatorio Astronómico Nacional (OAN), 28014 Madrid, Spain}
\author{David Navarro-Almaida}
\affiliation{Centro de Astrobiología (CAB), CSIC-INTA, Ctra. de Torrejón a Ajalvir, km 4, 28850 Torrejón de Ardoz, Spain}
\author{Marina Rodríguez Baras}
\affiliation{Observatorio Astronómico Nacional (OAN), 28014 Madrid, Spain}
\author{Angèle Taillard}
\affiliation{Centro de Astrobiología (CAB), CSIC-INTA, Ctra. de Torrejón a Ajalvir, km 4, 28850 Torrejón de Ardoz, Spain}
\author{Karin Öberg}
\affiliation{Center for Astrophysics | Harvard \& Smithsonian, Cambridge, MA 02138, USA}

\begin{abstract}
The isotopic ratios of deuterated methanol derived around protostars are commonly used to infer the physical conditions under which they formed in the earlier prestellar stage. However, there is a discrepancy in the ratio of the singly deuterated methanol isotopologues, CH$_2$DOH/CH$_3$OD, between low- and high-mass protostars, which puts into question whether prestellar isotopic ratios are generally preserved during the star- and planet-forming process. Resolving this puzzle is only made harder by the complete lack of data on this ratio in the prestellar stage. This work presents observations with the IRAM 30m telescope that securely detect CH$_3$OD in the prestellar core L1448 in Perseus and tentatively in B213-C6 in Taurus. This work constrains the ratio of CH$_2$DOH/CH$_3$OD and the D/H ratios for both singly deuterated methanol isotopologues for the first time at the prestellar stage. Column densities calculated under the assumption of local thermal equilibrium lead to a CH$_2$DOH/CH$_3$OD ratio of 2.8--8.5 in L1448 and $\leq$ 5.7 in B213-C6. The values are marginally consistent with the statistically expected ratio of 3, but most assumptions put the values in an elevated range in line with values found around low-mass protostars. The D/H ratio in CH$_2$DOH is between 3.6\% and 6.8\% in L1448 and in the range of 2.4--5.8\% in B213-C6. The D/H ratio derived for CH$_3$OD is lower, namely 1.4--4.4\% in L1448 and $\leq$ 3.8\% in B213-C6.
\end{abstract}


\section{Introduction} \label{sec:intro}
Starless and dynamically evolved prestellar cores represent the birthplaces of low-mass stars like our Sun (M$_*$ $<$ a few M$_\odot$). They are characterized by cold temperatures (T $\sim$~10~K) and gas densities $\geq$ 10$^4$~cm$^{-3}$ (\citealt{Benson89,Bergin07,Andre14}). Investigating their chemical composition allows us to learn about chemical processes at the initial stages of star formation, and lets us understand the initial molecular inventory that is inherited by subsequent stages of star and planet formation. 
\newline \indent Molecules are ubiquitous in space, and complex organic molecules (COMs; defined as carbon-bearing molecules with 6 or more atoms; \citealt{Herbst09}) have already been detected in the gas phase in prestellar cores (e.g., \citealt{Bacmann12,JimenezSerra16}). A key molecule to kickstart chemical complexity in space is methanol, CH$_3$OH (\citealt{Oberg09,Chuang16}), it is the simplest and one of the most widespread COMs and is detected across all stages of star formation, and even in Solar System comets (\citealt{Parise06,Bizzocchi14,Booth21,Drozdovskaya21}). Methanol predominately forms in CO-rich ices via the hydrogenation of CO, or reactions between the intermediates of this hydrogenation sequence at conditions found in prestellar cores (T $<$ 20~K, n$_{\rm H_2}$ $\geq$~10$^{5}$ cm$^{-3}$; \citealt{Watanabe02,Fuchs09,Santos22}). Due to its efficient formation in cold ice and lack of gas-phase formation routes, methanol is uniquely suited to infer the past conditions under which ices formed (\citealt{Geppert06}).\\
\indent A powerful tool to trace the historical physicochemical conditions during the star formation process is deuteration, a process that swaps out hydrogen for its heavier isotope, deuterium, and enriches molecules beyond the average interstellar D/H value of $\sim$~10$^{-5}$ (\citealt{Linsky06}). D/H ratios on the order of 10$^{-1}$--10$^{-4}$ are routinely obtained towards prestellar cores and protostars (\citealt{Nomura22}, \citealt{Esplugues22}, \citealt{Rodriguez-Baras23}). At temperatures below 20~K and gas densities $>$~10$^4$~cm$^{-3}$, enrichment in deuterium occurs via the reaction H$_3 ^+$ + HD $\rightleftharpoons$ H$_2$D$^+$ + H$_2$ + 232~K (\citealt{Watson74}). Once the temperature drops below 20~K, the backward reaction is quenched, and the ratio of H$_2$D$^+$/H$_3 ^+$ increases compared to the elemental D/H ratio. The atomic reservoir that is now enriched in deuterium can participate in grain-surface reactions resulting in isotopic enrichments in molecules like methanol once neutral species such as CO freeze out. Thus, the D/H ratio of methanol directly depends on the elevated D/H ratio in the gas phase (\citealt{Millar89}). Deuterated molecules are generally hard to detect in the solid state, only recently the first detection of a deuterated molecule in the ice, HDO, was claimed (\citealt{Slavicinska24}). Instead, we rely on gas-phase spectroscopy following ice sublimation to trace the level of deuterium fractionation in ice chemistry products (\citealt{Caselli12}). \\
\indent Both singly deuterated methanol isotopologues, CH$_2$DOH and CH$_3$OD, have been readily detected around protostars of all masses (e.g., \citealt{Parise06,Peng12,Neill13,Belloche16,Bögelund18,Jorgensen18}). At the prestellar stage, only CH$_2$DOH has been detected so far (\citealt{Bizzocchi14,Ambrose21,Lin23}). The D/H ratio of CH$_2$DOH in low-mass prestellar cores is consistent with ratios obtained in low-mass Class 0/I protostars and comet 67P/Churyumov-Gerasimenko (\citealt{Drozdovskaya21}), which suggests that their isotopic record is set at and subsequently inherited from the prestellar stage. H-D surface substitution that elevates D/H ratios in the CH$_3$-group of methanol (\citealt{Nagaoka05}) can explain the isotopic ratios of CH$_2$DOH, as shown in astrochemical models (e.g., \citealt{Taquet12,Kulterer22,Riedel23}). The D/H ratio in CH$_2$DOH around high-mass protostars is lower compared to their low-mass counterparts, which could be due to slightly higher temperatures during their time of formation (\citealt{vanGelder22}). This is also consistent with the D/H ratio of CH$_2$DOH in high-mass prestellar cores (\citealt{Fontani15}).\\
\indent There are some indicators; however, that the D/H ratio set at formation can be altered through ice chemical processes, veiling our extraction of ice formation environments from D/H ratios. The most pressing evidence is a claimed difference in D/H enhancement patterns in the OH-group of methanol between low- and high-mass protostars (\citealt{Taquet19}). While the low-mass pre- and protostellar fractionation patterns generally conform to the expectations from theoretical models, the fractionation ratios towards high-mass protostars do not. In addition, the ratio of CH$_2$DOH/CH$_3$OD deviates from its statistically expected value of 3: for low-mass protostars, this value can exceed 10, while it drops towards unity and below for high-mass protostars (e.g., \citealt{Parise06,Bögelund18}). It is not known which chemical process leads to the observed discrepancy in this ratio, and at which stage during the star formation process it takes place. In order to constrain whether this ratio is set at and inherited from the prestellar stage, or if it is affected by reprocessing of prestellar ices in the protostellar stage requires observations of CH$_3$OD at the prestellar stage. A similar ratio of CH$_2$DOH/CH$_3$OD at pre- and protostellar stages would suggest that it is already set in prestellar cores, while a deviation would indicate that chemical processes during the warm-up stage alter the ratio, either through selective formation or selective destruction of one of the isotopologues.  \\
\indent We present the first detection of CH$_3$OD in the low-mass prestellar core L1448 in the Perseus star-forming region, and an upper limit in B213-C6 in the Taurus star-forming region, which allows us to put a first constraint on the ratio of CH$_2$DOH/CH$_3$OD in prestellar cores, and assess if the protostellar fractionation pattern of methanol is reflected by hydrogen fractionation chemistry in prestellar cores in the low-mass case. The paper is structured as follows: we detail the observations in Section \ref{sect:observations}, and describe the analysis process in Section \ref{sect:analysis}. The detected singly deuterated methanol isotopologues and their isotopic ratios are described in Section \ref{sect:results}, and discussed comparatively with detections across the star-forming sequence in \ref{sect:discussion} to shed light on potential formation pathways resulting in the observed ratios. We raise our conclusions in Section \ref{sect:conclusion}.

\section{Observations}\label{sect:observations}
The prestellar core L1448 (RA = 03$^{\rm h}$25$^{\rm m}$49$^{\rm s}$.00, DEC = 30$^\circ$42$^{\prime}$24$^{\prime \prime}$.6) is located in the Perseus star-forming region; the prestellar core B213-C6 (RA = 04$^{\rm h}$18$^{\rm m}$08$^{\rm s}$.40, DEC = 28$^{\circ}$05$^{\prime}$12$^{\prime \prime}$) is located in Taurus. B213-C6 lies in the northern part of the B213 filament. This part is surrounded by $\sim$~40 protostars (\citealt{Luhman09,Rebull2010}). While the Perseus star-forming region is associated with two clusters in which pre-main sequence stars are currently forming (\citealt{Lada96,Luhman03}). L1448 is located in a quiescent part of the region that is not associated with active star formation. \\
\indent This work analyses data from two observational programs carried out at the IRAM 30m telescope. The first data set is the large program \textbf{G}as phase \textbf{E}lemental abundances in \textbf{M}olecular Cloud\textbf{S} (GEMS) and the second is project 071-23 (PI of both: A. Fuente). The GEMS observations were conducted in frequency-switching mode with a frequency throw of 6~MHz. By using the Eight MIxer Receivers (EMIR) in combination with the Fast Fourier Transform Spectrometers (FTS) as the backend; a spectral resolution of 49~kHz was achieved (\citealt{Fuente19}). This paper makes use of the CH$_2$DOH transitions at 85.2967, 86.6688, and 134.0658~GHz in the GEMS data. The rms for those transitions is between 13 and 20~mK. \\
\indent The second data set is project 071-23 with data taken on the 2$^{\rm nd}$ and 3$^{\rm rd}$ of May 2024 in excellent weather (pwv~=~0--4.6~mm, $\tau$~$<$~0.1) and at T$_{\rm sys}$ of 60--130~K. As for the GEMS data, EMIR with the FTS backend and a spectral resolution of 49~kHz was used. The data were also taken in frequency-switching mode with a frequency throw of 3.9~MHz. The four spectral windows were centered on 90.1, 93.6, 106.1, and 109.6~GHz, the rms for the spectral windows relevant for this work is 2.1--2.9~mK.\\
\indent The half power beam width (HPBW) is related to the frequency of the observed transition via\\ HPBW($^{\prime\prime}$)~=~2460/$\nu$\footnote{\url{https://publicwiki.iram.es/Iram30mEfficiencies}}, where $\nu$ is the frequency in GHz. This leads to a beam size of $\sim$~29$^{\prime\prime}$ at 85~GHz and a beam size of $\sim$~18$^{\prime\prime}$ at 134~GHz. The intensity scale of the data is the main beam temperature, T$_{\rm MB}$, which is related to the antenna temperature, T$_{\rm *,A}$ via T$_{\rm MB}$ = (F$_{\rm eff}$/B$_{\rm eff}$)~$\cdot$~T$_{\rm *, A}$. The beam efficiencies are listed in Table B.1 in \cite{Fuente19}. We note that CH$_2$DOH at 89.4078 and 90.7798~GHz is detected in both projects in both cores, but due to the higher noise in the GEMS data, combining data sets did not improve the signal-to-noise significantly, and we thus do not include the GEMS data of those two lines in our analysis.

\section{Analysis}\label{sect:analysis}
The initial line identification process was conducted with the \textsc{class} package of the \textsc{gildas}\footnote{\url{http://www.iram.fr/IRAMFR/GILDAS}} software, and the data were reduced with the automated \textsc{gildas-class} Pipeline (GCP\footnote{\url{https://github.com/andresmegias/gildas-class-python/ }}; \citealt{Megias23}). Line frequencies and the spectroscopic parameters of the transitions were taken from the Jet Propulsion Laboratory (JPL; \citealt{Pickett98}) for CH$_2$DOH and from the Cologne Database for Molecular Spectroscopy (CDMS; \citealt{Muller01,Muller05,Endres16}) for CH$_3$OD based on the spectroscopic entry of \cite{Ilyushin24}. The spectroscopic entry of CH$_2$DOH in JPL based on \cite{PEARSON2012119} states that column densities calculated from b- and c-type transitions are not reliable. We therefore only used the two a-type transitions at 89.4078 and 90.7798~GHz for determining the column density of CH$_2$DOH. Partition function values below 10~K for CH$_3$OD are not publicly available, so we calculated the partition function values at 6 and 8~K from the energy levels of the $v_{t} = 0$ torsion level (H. Müller, private communication). For CH$_2$DOH, we used an updated catalogue to obtain the partition function values (L. Coudert, private communication). The respective partition function values are listed in Table \ref{tab:pfval}. The peak temperature of each detected line, T$_{\rm MB}$ (K), alongside the center velocity, v$_{\rm 0}$ (km/s), and the line width, $\delta$v (km/s), of each detected line was determined by fitting a Gaussian line profile utilizing the \textsc{curve\_fit} package of \textsc{scipy} (\citealt{scipy}). The resulting values are shown in Table \ref{tab:detectedlines}. \\
\indent The calculation of the column densities of CH$_2$DOH and CH$_3$OD was carried out under the assumption of local thermal equilibrium (LTE). For cores L1448 and B213-C6, it has been shown by \cite{Spezzano22} and Kulterer et al. (submitted) that CH$_3$OH cannot be fit using an LTE approach, but because collisional data for the deuterated methanol isotopologues are not available, we are left with using the LTE approximation. In the case of LTE, the line flux $\int$T$_{\rm MB}$dv (K/km/s) is related to the column density in the upper state (N$_{\rm up}$ (cm$^{-2}$)) via
\begin{equation}\label{eq:Nup}
    \rm N_{up} = \int T_{MB} dv \times \frac{8\pi \cdot k_B \cdot \nu^2}{h \cdot c^3 \cdot A_{ul}},
\end{equation}
where k$_{\rm B}$ corresponds to the Boltzmann constant, $\nu$ is the frequency of the transition, h is the Plank constant, c the speed of light, and A$_{\rm ul}$ the Einstein A coefficient of the upper energy level. The peak temperature T$_{\rm MB}$ (K) can be calculated from
\begin{equation}\label{eq:TMB}
    \rm T_{MB} = \frac{\int T_{MB}dv}{FWHM} \times \frac{2\sqrt{2ln(2)}}{\sqrt{2\pi}},
\end{equation}
with the FWHM being the full width half maximum ($\delta$v) of the line. N$_{\rm up}$, the column density in the upper energy state, is determined via 
\begin{equation}
    \rm N_{up} = \frac{g_{up} \cdot N_{tot}}{Q(T_{ex}) \cdot exp(E_{up}/k_B T_{ex})}.
\end{equation}
Q(T$_{\rm ex}$) corresponds to the partition function at an excitation temperature T$_{\rm ex}$, E$_{\rm up}$ and g$_{\rm up}$ correspond to the energy and the degeneracy of the upper state, respectively. 
As a first step, we attempted to fit one column density and excitation temperature if multiple lines of one molecule were detected with the \textsc{cassis} software\footnote{\url{http://cassis.irap.omp.eu/}} (\citealt{Vastel15}). We explored a grid of T$_{\rm ex}$ from 10 to 15~K in steps of 0.1~K and N$_{\rm tot}$ values of (5~$\times$~10$^{11}$)--(5~$\times$~10$^{13})$~cm$^{-2}$ with a step size of 0.01 in logarithmic space. The source size was set to 27$^{\prime\prime}$, which equals the beam size at 90~GHz, and we assumed an average line width from the detected transitions for each molecule. However, it was not possible to determine a good fit: the lowest $\chi ^2$ obtained was $\sim$~10, which may be due to non-LTE effects that are not included in the calculations. On the other hand, the temperature range was set to mimic the dust temperature range derived for L1448 and B213-C6 by \cite{RodriguezBaras21}; and the gas could potentially be colder. As a second approach, we instead calculated the best-fit column density for each transition for fixed T$_{\rm ex}$ of 6, 8, and 10~K from the integrated intensity $\int T_{\rm MB}$~dv. The column density for a molecule is calculated from the average of the individual best-fit column densities for each line. The column densities for the assumed grid of excitation temperatures are listed in Table \ref{tab:CDDH}.

\section{Results}\label{sect:results}

\subsection{Detected Molecules}\label{sect:detectedmol}
\begin{figure*}
    \includegraphics[width=1.0\textwidth]{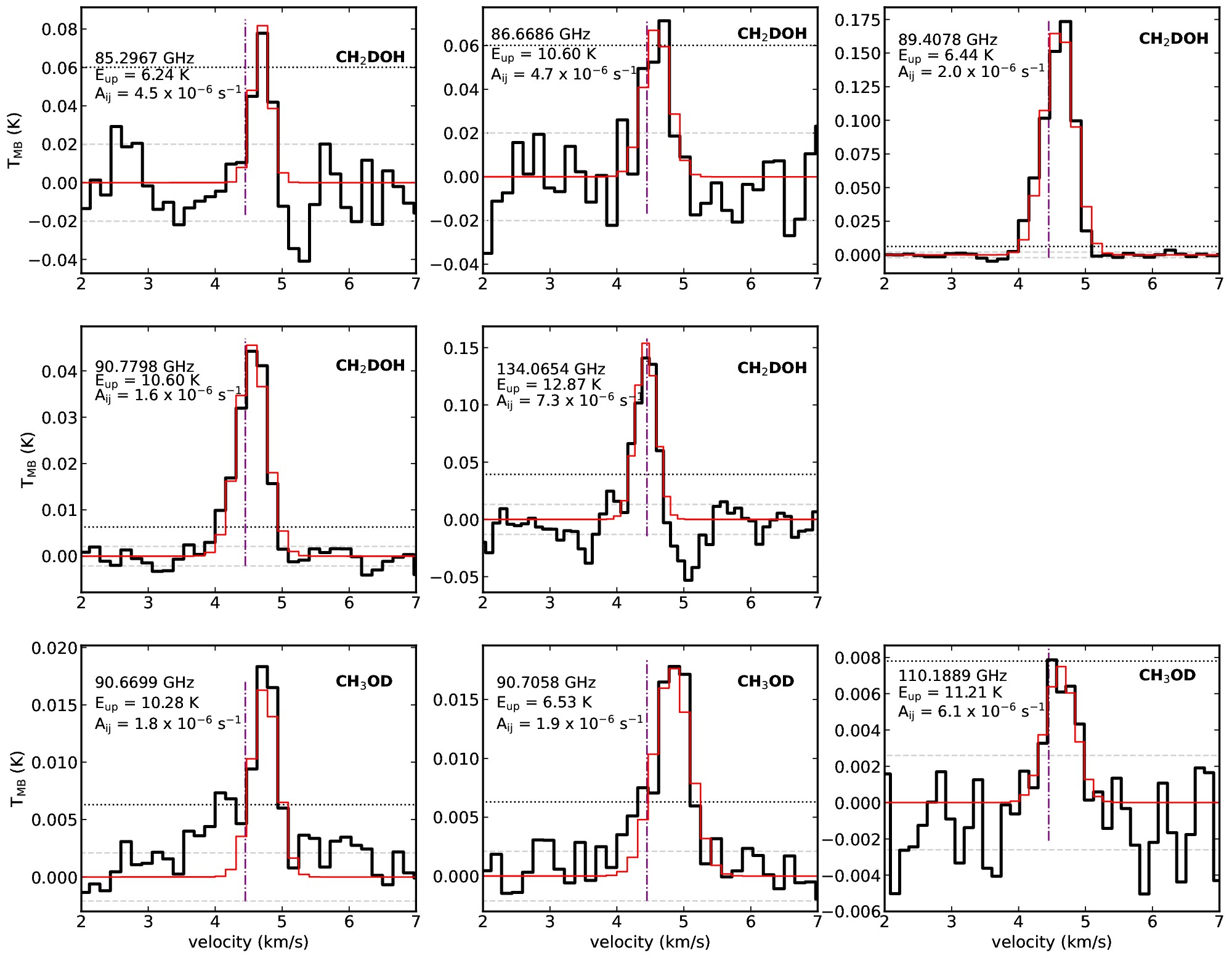}
    \caption{Detected transitions of CH$_2$DOH (top two rows) and CH$_3$OD (bottom row) towards L1448. The line at 110~GHz is a tentative detection. The observed spectra plotted in black are overlaid by the Gaussian fits (see Table \ref{tab:detectedlines}) in red. The silver dashed lines represent the value of $\pm$1 $\times$ rms, and the dotted, black line represents the value of 3 $\times$ rms. The purple dash-dotted vertical line represents the line rest velocity of CH$_3$OH (Kulterer et al. submitted).}
    \label{fig:L1448_Spec}
\end{figure*}

\begin{figure*}
    \includegraphics[width=1.0\textwidth]{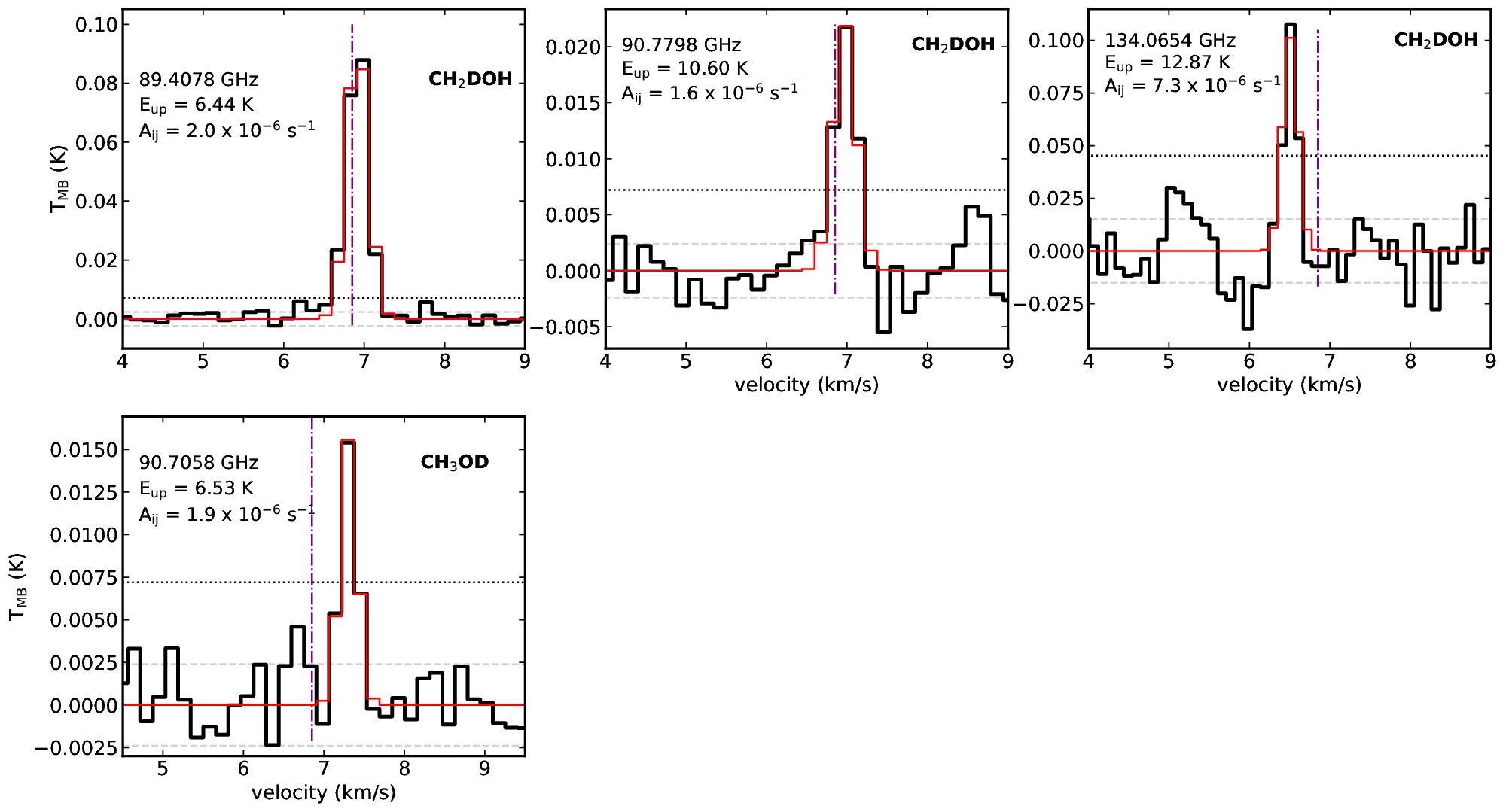}
    \caption{Detected transitions of CH$_2$DOH (top row) and CH$_3$OD (bottom left panel) towards B213-C6. The observed spectra plotted in black are overlaid by the Gaussian fits (see Table \ref{tab:detectedlines}) in red. The silver dashed lines represent the value of $\pm$1 $\times$ rms, and the dotted, black line represents the value of 3 $\times$ rms. The purple dash-dotted vertical line represents the line rest velocity of CH$_3$OH (\citealt{Spezzano22}).}
    \label{fig:B213C6_Spec}
\end{figure*}

For the first time, CH$_3$OD has been detected in a prestellar core. Its transitions at 90.6699, 90.7058, and tentatively at 110.1889~GHz are detected in L1448 (Fig. \ref{fig:L1448_Spec}). We consider the detection at 110.1889~GHz as tentative, because the peak intensity of the line equals only 3~$\times$~rms. A line at 90.7058~GHz is detected towards B213-C6 (Fig. \ref{fig:B213C6_Spec}), which matches the position of a CH$_3$OD transition and within 1~MHz of this line, there is no other molecule identified in the online tool Splatalogue\footnote{\url{https://splatalogue.online}} with transitions at upper level energies below 30~K. However, we only find this one line in the spectra of B213-C6, and therefore consider this only a tentative indication of CH$_3$OD in this core and treat the column densities and ratios of CH$_3$OD in B213-C6 as limits. At the position of the 90.6699 GHz transition in B213-C6 a signal just slightly above the noise level is detected with an intensity of 2.4~mK. In addition, we report the detections of five CH$_2$DOH lines towards L1448 and three lines towards B213-C6. \\
The respective frequencies, upper level energies, Einstein A coefficients of the lines, as well as their fitted peak intensities, integrated intensities, line widths, and center velocities are listed in Table \ref{tab:detectedlines}. The center velocities of the lines deviate by less than 0.5~km/s from the main methanol isotopologue: v$_{\rm 0}$ for CH$_3$OH in L1448 is 4.45~km/s (Kulterer et al. submitted) and 6.85~km/s in B213-C6 (\citealt{Spezzano22}). The line widths of the deuterated isotopologues are slightly different from that of CH$_{3}$OH in L1448 ($\delta$v$_{\rm CH_3 OH}$ = 0.57~km/s; Kulterer et al. submitted). {The majority of the CH$_2$DOH lines are 5--20~\% narrower, while the CH$_3$OD lines are within 20\% of the line width of CH$_3$OH. Within errors, the lines of the isotopologues lie within 2-3 channels at most from each other, so they emit from nearly the same region. For B213-C6, the lines of the deuterated isotopologues are on average only about half the line width of the main isotopologue ($\delta$v$_{\rm CH_3 OH}$ = 0.53~km/s; \citealt{Spezzano22}), which hints that CH$_3$OH in B213-C6 detected in the data might stem from multiple gas layers, while deuterated methanol emits from only one region. It is to note that the peak of the suspected CH$_3$OD transition at 90.7058~GHz in B213-C6 differs by four channels ($\sim$ 0.38~km/s) from the velocity of CH$_3$OH, while the CH$_2$DOH lines are within two channels. \\
\begin{table*}[h]
    \caption{Observed transitions of CH$_2$DOH and CH$_3$OD in L1448 and B213-C6. Tentative detections are marked with an asterisk in front of the transition frequency. The plots of the spectra and their fits are shown in Figs. \ref{fig:L1448_Spec} and \ref{fig:B213C6_Spec}.}\label{tab:detectedlines}
    \centering
    \begin{tabular}{lllrrcrrccr}
    \hline \hline 
    Core & Molecule & Transition & Frequency & E$_{\rm up}$ & A$_{\rm ij}$ & T$_{\rm mb, peak}$ & $\int T_{\rm MB}$ ~dv & $\delta$v & v$_{\rm 0}$ & T$_{\rm rms}$ \\
    & &  & (GHz) & (K) & (10$^{-6}$ s$^{-1}$) & (K) & (K km s$^{-1}$) & (km~s$^{-1}$) & (km~s$^{-1}$) & (mK) \\
    \hline 
    L1448 & CH$_2$DOH & 1$_{\rm 1,0}$ - 1$_{\rm 0,1}$ & 85.29673 & 6.24 & 4.5 & 82.1 $\pm$ 18.5 & 28.4 $\pm$ 9.2 & 0.33 $\pm$ 0.08 & 4.62 $\pm$ 0.03 & 20.0 \\
    & & 2$_{\rm 1,1}$ - 2$_{\rm 0,2}$ & 86.66875 & 10.60 & 4.7 & 69.0 $\pm$ 12.5 & 34.5 $\pm$ 9.4 & 0.47 $\pm$ 0.10 & 4.52 $\pm$ 0.04 & 20.0 \\
    & & 2$_{\rm 0,2}$ - 1$_{\rm 0,1}$ & 89.40782 & 6.44 & 2.0 & 171.2 $\pm$ 3.8 & 98.2 $\pm$ 3.9 & 0.54 $\pm$ 0.01 & 4.53 $\pm$ 0.01 & 2.1 \\
    & &2$_{\rm 1,1}$ - 1$_{\rm 1,0}$ & 90.77984 & 6.24 & 1.6 & 45.6 $\pm$ 1.6 & 25.5 $\pm$ 1.4 & 0.53 $\pm$ 0.02 & 4.48 $\pm$ 0.01 & 2.1 \\
    & & 3$_{\rm 0,3}$ - 2$_{\rm 0,2}$ & 134.06583 & 6.44 & 7.3 & 154.1 $\pm$ 2.0 & 59.5 $\pm$ 5.4 & 0.36 $\pm$ 0.03 & 4.38 $\pm$ 0.01 & 13.2 \\
    & CH$_3$OD & 2$_{\rm 0,0}$ - 1$_{\rm 0,0}$~$^{+}$ & 90.66987 & 10.28 & 1.8 & 16.6 $\pm$ 2.0 & 8.3 $\pm$ 1.5 & 0.47 $\pm$ 0.06 & 4.66 $\pm$ 0.04 & 2.1 \\
    & & 2$_{\rm 0,0}$ - 1$_{\rm 0,0}$~$^{-}$& 90.70577 & 6.534 & 1.9  & 17.8 $\pm$ 1.3 & 12.0 $\pm$ 1.4 & 0.64 $\pm$ 0.06 & 4.75 $\pm$ 0.02 & 2.1 \\
    & & 1$_{\rm 0,0}$ - 1$_{\rm 0,0}$~$^{+}$ & \textbf{*}110.18887 & 11.21 & 6.1  & 7.5 $\pm$ 1.8 & 4.2 $\pm$ 1.5 & 0.53 $\pm$ 0.14 & 4.56 $\pm$ 0.06 & 2.6 \\
    B213-C6 & CH$_2$DOH & 2$_{\rm 0,2}$ - 1$_{\rm 0,1}$ & 89.40782 & 6.44 & 2.0 & 96.3 $\pm$ 3.2 & 32.9 $\pm$ 1.4 & 0.32 $\pm$ 0.01 & 6.84 $\pm$ 0.01 & 2.4 \\
    & & 2$_{\rm 1,1}$ - 1$_{\rm 1,0}$ & 90.77984 & 6.24 & 1.6  & 21.9 $\pm$ 2.9 & 8.0 $\pm$ 1.6 & 0.34 $\pm$ 0.05 & 6.90 $\pm$ 0.02 & 2.4 \\
    & & 3$_{\rm 0,3}$ - 2$_{\rm 0,2}$ & 134.06583 & 6.44 & 7.3 & 101.2 $\pm$ 19.0 & 25.4 $\pm$ 7.2 & 0.24 $\pm$ 0.05 & 6.45 $\pm$ 0.02 & 15.1 \\
    & CH$_3$OD & 2$_{\rm 0,0}$ - 1$_{\rm 0,0}$~$^{-}$ & 90.70577 & 6.534 & 1.9  & 15.6 $\pm$ 1.8 & 4.4 $\pm$ 0.8 & 0.26 $\pm$ 0.04 & 7.23 $\pm$ 0.02 & 2.4 \\
    \hline 
    \end{tabular}
    \label{tab:my_label}
\end{table*}
Following the approach in Section \ref{sect:analysis}, the column density of CH$_2$DOH in L1448 is in the range of (0.9--1.5)~$\times$~10$^{13}$~cm$^{-2}$ and for CH$_3$OD in the range of (1.2--3.5)~$\times$~10$^{12}$~cm$^{-2}$ for excitation temperatures of 6--10~K (Table \ref{tab:CDDH}). In B213-C6, the column density of CH$_2$DOH is in the range of (2.5--5.9)~$\times$~10$^{12}$~cm$^{-2}$ and $\leq$ 1.3 ~$\times$~10$^{12}$~cm$^{-2}$ for CH$_3$OD. 
\subsubsection{Potential Caveats}
The line shape and intensity of the CH$_3$OD transition at 90.6699~GHz may be affected by the use of a frequency throw of 3.9~MHz during frequency switching. There is a nearby HNC line at 90.6636~GHz; and the artifact produced by the automated data reduction pipeline of the IRAM 30m telescope during the folding of the data is close to the CH$_3$OD transition (Fig. \ref{fig:freqthrow}). It is not possible to say if potential wiggles at the wing of the feature either increase or decrease the intensity of the line. The peak intensity of the CH$_3$OD transition is $>$~6~$\times$~rms in L1448, so we consider this line detected, but we doubled the errors in the integrated intensity that we used for the fitting of the column densities compared to the value listed in Table \ref{tab:detectedlines}. The peak intensity at the position of this transition in B213-C6 is only at the noise level. This non-detection can be attributed to non-LTE effects, because in LTE, the column density derived from the detected 90.7058 GHz transition should have led to a detection of this line too (Fig. \ref{fig:freqthrow2}).\\
\indent Another caveat that has to be considered is that the calculation of the column densities has been carried out based on the assumption that deuterated methanol is in LTE. This might not be the case. However, collisional data for CH$_2$DOH and CH$_3$OD are not available. A hint that deuterated methanol could be in non-LTE comes from the two a-type CH$_2$DOH transitions at 89.4078 and 90.7798~GHz. In LTE, their line intensity ratio should be $\sim$~2, but it is 4--5 in L1448 and B213-C6. In addition, given the sensitivity of the observations, the CH$_3$OD transitions at 90.6699 and 110.1889~GHz in B213-C6 should have just been detected at 3$\sigma$ for T$_{\rm ex}$ of 6--10~K based on the upper limit on the column density derived from the transition at 90.7058~GHz if CH$_3$OD would be in LTE.

\subsection{Isotopic Ratios of Singly Deuterated Methanol}\label{sect:DMeth_ratios}
In general, the column densities of the deuterated methanol isotopologues are one to two orders of magnitude less abundant than the main isotopologue. The column density of CH$_3$OH is (3.5 $\pm$ 0.4) $\times$ 10$^{13}$ cm$^{-2}$ in B213-C6 (\citealt{Spezzano22}) and (8.2 $\pm$ 0.7) $\times$ 10$^{13}$ cm$^{-2}$ in L1448 (Kulterer et al. submitted). In order to get from column densities to the D/H ratios of the two functional groups in methanol, one has to take statistical weighting into account, as a deuterium atom is three times more likely to land in the CH$_3$-group than the OH-group. The D/H ratio of singly deuterated methanol can therefore be calculated via 
\begin{equation}\label{eq:DH_CH2DOH}
    \rm N_{CH_2 DOH} / N_{CH_3OH} = 3(D/H)_{CH_3OH}
\end{equation}
and
\begin{equation}\label{eq:DH_CH3OD}
    \rm N_{CH_3 OD} / N_{CH_3OH} = (D/H)_{CH_3OH}.
\end{equation}
We calculated the D/H ratios of CH$_2$DOH and CH$_3$OD under the assumption that they emit from the same region as the main isotopologue, and present the values derived for excitation temperatures of 6, 8, and 10~K for deuterated methanol in Table \ref{tab:CDDH}. Depending on the assumed excitation temperature, the D/H ratio derived from CH$_2$DOH in L1448 is between 3.6\% and 6.8\%, while in B213-C6, it is found to range from 2.4\% to 5.8\%. The highest excitation temperature leads to the highest D/H ratio; and the ratio decreases with decreasing excitation temperature. The D/H ratio in CH$_3$OD is found to be between 1.4\% and 4.4\% in L1448, and $\leq$ 3.8\% in B213-C6. Assuming that the excitation temperature of CH$_2$DOH and CH$_3$OD is the same, the D/H ratio in the CH$_3$-group is always higher than in the OH-group. In the case of L1448, it is in the range of 40--50\%, where the lower percentage corresponds to the lowest excitation temperature. For B213-C6, the D/H ratio in CH$_2$DOH is $\sim$~33\% more elevated regardless of the excitation temperature. Furthermore, the deuterium fraction in CH$_2$DOH is $\sim$~30\% higher in L1448 than in B213-C6, independent of the choice of excitation temperature.\\
\indent Again, we assume that CH$_2$DOH and CH$_3$OD emit from the same region in order to calculate the ratio of CH$_2$DOH/CH$_3$OD. Based on statistical weighting, the ratio is expected to be 3, but most assumptions put the obtained values above the expected value. Depending on T$_{\rm ex}$, the ratio is between 2.8 and 8.5 in L1448, $\leq$ 5.7 in B213-C6. At 6~K, the ratio is 4.9 $\pm$ 2.1 in L1448 and $\leq$ 5.6 in B213-C6; at 8 K, it is 5.5 $\pm$ 2.3 in L1448 and $\leq$ 5.6 in B213-C6; and the highest values are found for 10~K, namely 6.0 $\pm$ 2.5 in L1448 and $\leq$ 5.7 in B213-C6. Only one line of CH$_3$OD has been detected in B213-C6, hence the ratios are treated as limits for this core.

\begin{table*}[]
    \caption{Column densities of both singly deuterated methanol isotopologues for T$_{\rm ex}$ of 6, 8, and 10 K, and D/H ratios of CH$_2$DOH and CH$_3$OD for T$_{\rm ex}$ = 6, 8, and 10 K relative to CH$_3$OH. We assume a column density for CH$_3$OH of (8.18 $\pm$ 0.70) $\times$ 10$^{13}$ cm$^{-2}$ (Kulterer et al. submitted) for L1448 and a column density of (3.50 $\pm$ 0.4) $\times$ 10$^{13}$ cm$^{-2}$ for B213-C6 (\citealt{Spezzano22}). In the last column, N$_{\rm corr}$ corresponds to the D/H ratio of CH$_2$DOH where statistical weighting has been taken into account. No statistical weighting is required for the D/H ratio of CH$_3$OD.}\label{tab:CDDH}
\centering
    \begin{tabular}{llllll}
    \hline \hline 
    Core & Molecule  & T$_{\rm ex}$ (K) & \textit{N$_{\rm tot}$} (cm$^{-2}$) &  \textit{N}/\textit{N}$_{\rm CH_3 OH}$ &   \textit{N}$_{\rm corr}$/\textit{N}$_{\rm CH_3 OH}$ \\
    \hline 
    L1448 & CH$_2$DOH & 6 & 9.84 $\pm$ 0.61 $\times$~10$^{12} $&  12.0 $\pm$ 1.3 \% & 4.0 $\pm$ 0.4 \% \\
    & & 8 & 1.20 $\pm$ 0.10 $\times$ 10$^{13}$ & 14.6 $\pm$ 1.6 \% & 4.9 $\pm$ 0.5 \% \\
    & & 10 & 1.50 $\pm$ 0.10 $\times$ 10$^{13}$ & 18.4 $\pm$ 2.0 \% & 6.1 $\pm$ 0.7 \% \\
    & CH$_3$OD & 6 & 1.96 $\pm$ 0.81 $\times$ 10$^{12}$ & 2.4 $\pm$ 1.0 \% & 2.4 $\pm$ 1.0 \% \\
    & & 8 & 2.17 $\pm$ 0.90 $\times$ 10$^{12}$ & 2.7 $\pm$ 1.1 \% & 2.7 $\pm$ 1.1 \% \\
    & & 10 & 2.50 $\pm$ 1.04 $\times$ 10$^{12}$ & 3.1 $\pm$ 1.3 \% & 3.1 $\pm$ 1.3 \% \\
    B213-C6 & CH$_2$DOH & 6 & 3.20 $\pm$ 0.67 $\times$ 10$^{12}$ & 9.1 $\pm$ 2.2 \% & 3.1 $\pm$ 0.7 \% \\
    & & 8 & 3.90 $\pm$ 0.82 $\times$ 10$^{12}$ & 11.2 $\pm$ 2.7 \% & 3.7 $\pm$ 0.9 \% \\
    && 10 & 4.92 $\pm$ 1.03 $\times$ 10$^{12}$ & 14.1 $\pm$ 3.4 \% & 4.7 $\pm$ 1.1\% \\
    & CH$_3$OD & 6 & 7.19 $\pm$ 1.09 $\times$ 10$^{11}$ & 2.1 $\pm$ 0.4 \% & 2.1 $\pm$ 0.4 \% \\
    & & 8 & 8.85 $\pm$ 1.57 $\times$ 10$^{11}$ & 2.5 $\pm$ 0.5 \% & 2.5 $\pm$ 0.5 \% \\
    & & 10 & 1.09 $\pm$ 0.19 $\times$ 10$^{12}$ & 3.1 $\pm$ 0.7 \% & 3.1 $\pm$ 0.7 \% \\
    \hline 
    \end{tabular}
\end{table*}


\section{Discussion}\label{sect:discussion}

\subsection{Origins of Prestellar Methanol Deuterium Fractionation Patterns}\label{sect:DMeth_fm}
The average D/H ratio in the  ISM is $\sim$~10$^{-5}$ (\citealt{Linsky06}). However, the D/H ratios in CH$_2$DOH and CH$_3$OD are elevated beyond this value in prestellar cores, which shows that processes leading to their efficient deuteration must be present in prestellar cores. \\
\indent Methanol can form through successive hydrogenation of CO and from methane oxidation in ices (\citealt{Watanabe02,Qasim18,Santos22}) under conditions that are found in prestellar cores. In the case of methanol formation starting from CO, deuterium fractionation can occur through at least three mechanisms. One possibility to form CH$_2$DOH and CH$_3$OD in prestellar ices is by replacing one H atom with one D atom in the hydrogenation sequence 
\begin{equation}
    \rm CO \rightarrow HCO \rightarrow H_2 CO \rightarrow CH_3 O, CH_2 OH \rightarrow CH_3 OH,
\end{equation}
when temperatures below 20~K and gas densities at $\geq$ 10$^4$ cm$^{-3}$ (\citealt{Watanabe02,Caselli12}) lead to freeze-out of CO from the gas and enhanced availability of D atoms in the gas. This mechanism has been tested by models in \cite{Kulterer22}, and the models found that ratios of CH$_2$DOH/CH$_3$OD beyond the statistically expected value of 3 are only found for models with dust temperatures of 10~K, core ages $>$~3~$\times$~10$^5$~yr, and gas densities $\geq$~10$^6$~cm$^{-3}$. While this mechanism surely contributes to the abundance of deuterated methanol, it is not sufficient to explain the ratios of CH$_2$DOH/CH$_3$OD in low-mass prestellar cores and protostars (e.g. \citealt{Taquet12,Kulterer22,Riedel23}). \\
\indent In addition to the hydrogenation pathway, methanol can also form directly via
\begin{equation}\label{eq:Santos}
    \rm CH_3 O + H_2 CO \rightarrow CH_3 OH + HCO.
\end{equation}
Experiments by \cite{Santos22} have shown that this is the dominant formation pathway for methanol at temperatures of 10--16~K. These experiments have also shown that deuteration in this formation pathway occurs preferentially in the CH$_3$-group, if H$_2$CO is replaced by HDCO in Eq. \ref{eq:Santos}. Thus, this formation mechanism is another means to a) increase the D/H ratio in CH$_2$DOH, and b) boost the ratio of CH$_2$DOH/CH$_3$OD at the prestellar stage, but it remains to be tested in astrochemical models.\\
\indent Another formation pathway that leads to efficient and successive deuteration of the methyl group of methanol is abstraction. Here, an impinging H or D atom removes an H atom from the methyl group, e.g., CH$_3$OH + H $\rightarrow$ CH$_2$OH + H (\citealt{Nagaoka05}). Depending on whether CH$_2$OH reacts with H or D afterward, it either forms CH$_3$OH again or forms CH$_2$DOH. Furthermore, CH$_2$DOH can get successively deuterated, if abstraction occurs again, experiments have shown that once incorporated in the molecule, D atoms do not get abstracted anymore (\citealt{Nagaoka05,Hidaka09}). This mechanism can thus lead to a) a ratio of CH$_2$DOH/CH$_3$OD higher than the statistically expected value of 3, and b) high D/H ratios in CH$_2$DOH compared to CH$_3$OD. Indeed, models by \cite{Kulterer22} have shown that implementing this formation pathway is crucial for elevating the ratio of CH$_2$DOH/CH$_3$OD beyond the statistically expected value of 3 at prestellar stages. Moreover, \cite{Riedel23} have implemented this abstraction scheme in a source-tailored model for the prestellar core L1544 and are able to closely reproduce the observed column densities of CH$_3$OH and CH$_2$DOH in L1544. In addition, this mechanism should lead to D/H ratios in CHD$_2$OH and CD$_3$OH that should be higher than in CH$_2$DOH. This has indeed been confirmed by observations (e.g., \citealt{Drozdovskaya22,Ilyushin23,Lin23,Scibelli25}). This mechanism can thus explain the high D/H ratios in CH$_2$DOH. In line with experimental work (\citealt{Nagaoka05,Hidaka09}), the lower dust temperatures in low-mass star-forming regions favor this process, higher temperatures in high-mass star-forming regions quench its efficiency, leading to lower D/H ratios in CH$_2$DOH for high-mass objects, and a similar abundance of CH$_2$DOH and CH$_3$OD. This can explain the trend that we see in CH$_2$DOH in Fig. \ref{fig:DHratios} in low-mass prestellar cores compared to high-mass prestellar cores. \\
\indent Methanol formation can also occur with methane as a starting point via two routes. Experiments by \cite{Qasim18} have shown that CH$_3$OH can form in H$_2$O-rich ices before CO freezes out. In such a scenario, methanol forms from
\begin{equation}\label{eq:Qasim1}
    \rm CH_4 + OH \rightarrow CH_3 + H_2 O,
\end{equation}
followed by 
\begin{equation}\label{eq:Qasim2}
    \rm CH_3 + OH \rightarrow CH_3 OH.
\end{equation}
Deuterated methanol can form if deuterated methane or OD are present. However, it is not known, if the D atom in CH$_3$D prefers to go into CH$_2$D or into HDO in Eq. \ref{eq:Qasim1}; and thus, if it even has the potential to form deuterated methanol.\\
\indent Lastly, methanol can also form upon insertion of excited O($^1$D) into CH$_4$, where excited O($^1$D) forms after O$_2$ gets photodissociated into O($^3$P) + O($^1$D) (\citealt{Bergner17}). Similar to the formation pathway investigated by \cite{Qasim18}, this pathway could also occur starting from CH$_3$D, but it is again unknown if this preferentially forms CH$_2$DOH or CH$_3$OD. Whether the formation of the singly deuterated methanol isotopologs behaves statistically, or if the difference in energies required for breaking a C-H vs. a C-D bond is high enough to make a difference has to be investigated by either further experiments or computational studies. Only then it will be possible to assess how this formation pathway influences the ratio of CH$_2$DOH/CH$_3$OD and the D/H ratio of the methanol isotopologues in prestellar ices.\\
\indent Our current knowledge from experiments and observations suggests that the abstraction scheme can explain the observed D/H patterns in CH$_2$DOH best. The bulk of CH$_3$OD is likely formed via atom addition to CO, whether the routes with methane as a starting point preferentially form CH$_3$OD and thus, change the ratio of CH$_2$DOH/CH$_3$OD, is yet to be investigated. The D/H ratios from low-mass prestellar cores translate well to protostars (Fig. \ref{eq:DH_CH2DOH}), which suggests that the majority of deuterated methanol indeed stems from the prestellar stage, at least in the low-mass case.

\subsection{Survival of Prestellar D/H Ratios in Methanol}
The observed D/H levels in methanol across star-forming stages, source types, and isotopologues are not consistent. As seen in Fig. \ref{fig:DHratios}, the D/H fractionation in CH$_2$DOH in low-mass prestellar cores and low-mass protostars is higher than in CH$_3$OD. On the other hand, the average D/H fractionation is similar in CH$_2$DOH and CH$_3$OD for high-mass protostars. Consequently, the ratio of CH$_2$DOH/CH$_3$OD is higher for the low-mass case than the high-mass case (Fig. \ref{fig:CH2DOH_CH3OD}). No measurement of CH$_3$OD exists in high-mass prestellar cores, so it is not possible to verify if this trend also extends to high-mass prestellar sources.\\
\indent However, independent of the formation stage or environment, all D/H ratios are higher by at least one order of magnitude compared to the ISM average of $\sim$~10$^{-5}$ (\citealt{Linsky06}). The here-presented D/H ratios of methanol and the CH$_2$DOH/CH$_3$OD ratio in prestellar cores strongly suggest that the bulk of deuterated methanol is formed at the prestellar stage (Fig. \ref{fig:DHratios}, \ref{fig:CH2DOH_CH3OD}), but that there might be processes that alter the inherited isotopic fingerprint at subsequent stages.\\
\indent One such mechanism is the H-D exchange between the OH-group of methanol and water, which can alter the abundances of the main methanol isotopologue and CH$_3$OD. This isotope exchange occurs at temperatures as low as 120~K (\citealt{Ratacjzak09}). H-D exchange on the methyl group of methanol has been found to be inefficient (\citealt{Osamura04}), and thus, only the abundance of hydroxyl-deuterated methanol is going to be influenced by this process. For instance, this mechanism has been used to explain the column densities of CH$_3$OD in Orion KL (\citealt{Wilkins22}). However, this thermal exchange of H and D between methanol and water has not only the potential to form CH$_3$OD, it can also destroy it to form CH$_3$OH. Whether formation or destruction of CH$_3$OD is dominating depends on the reaction dynamics, and the starting ratios of D/H in water and methanol. In general, the likelihood of a CH$_3$OD molecule meeting a H$_2$O molecule in the ice is a lot higher than CH$_3$OH encountering HDO or D$_2$O. H$_2$O is the most abundant ice species, while the abundance of the main methanol isotopologue is a few percent w.r.t. to water (\citealt{Boogert15}). It is thus likely that this mechanism leads to changes in the D/H ratio of prestellar CH$_3$OD, but the extent of this alteration is unknown. \\
\indent Once sublimated from the grains, the abundance of CH$_3$OD can be altered via ion-molecule reactions in the gas phase (\citealt{Osamura04}) via
\begin{equation}
    \rm CH_3 OD + X-H^+ \rightarrow CH_3 OHD^+ + X,
\end{equation}
where X-H$^+$ is a protonating ion, followed by the reaction
\begin{equation}
    \rm CH_3 OHD^+ + e^- \rightarrow CH_3 OH + D. 
\end{equation}
\indent To which extent this could change the inherited isotopic ratios of methanol has not been tested in models. The discussed mechanisms suggest that the D/H ratios in methanol could be mainly set at the prestellar stage and potentially marginally altered during the protostellar stage. To verify this hypothesis, we calculated D/H ratios of methanol and ratios of CH$_2$DOH/CH$_3$OD by compiling CH$_2$DOH and CH$_3$OD detections from the literature, which are presented in the Appendix in Table \ref{tab:DHreferences}. The table has been adapted from \citealt{Drozdovskaya21}, such that observations toward objects from this previous compilation are included if, both, singly deuterated isotopologues are detected or an upper limit for CH$_3$OD has been derived. If only upper limits of CH$_2$DOH are available for a source, it has been excluded. In addition, it has been updated with detections since the publication of that paper. \\
\indent The average D/H ratio in CH$_2$DOH between low-mass prestellar cores and low-mass protostars differs by less than a factor of 2 (left panel Fig. \ref{fig:DHratios}). For low-mass prestellar cores, the values are in the range of $\sim$ 1--10\%. The spread of values is larger for the protostellar evolutionary stage, with values in the range of $\sim$~0.2--30\% including the most extreme cases. It is to note that all prestellar core values have been determined from single-dish facilities, while some of the protostellar data have been observed with interferometers, which leads to the possibility to derive the abundances from emission at smaller spatial scales. Single-dish observations of protostars capture emission of the cold envelope, where material is less affected by the presence of the central heating source, and the D/H ratios are most likely still pristine. The temperature structure around protostars is less homogeneous than in prestellar cores, so the collected data stems from different emitting regions, which could explain the larger spread in ratios observed around protostars. The overall high D/H ratios favor a prestellar origin; and the current data strongly suggest that the isotopic fingerprints from the two stages in CH$_2$DOH are connected. \\
\indent The average D/H ratio in CH$_2$DOH does not change from high-mass prestellar cores to high-mass protostars, but, again, the spread in values is larger for the protostellar case. High-mass protostars display D/H ratios between 10$^{-4}$ and 10$^{-1}$; in high-mass prestellar cores, the values are in the range of $\sim$~0.05--2\%.\\
\indent In general, while the D/H ratios in CH$_2$DOH are consistent for low-mass prestellar cores and low-mass protostars, and for high-mass prestellar cores and high-mass protostars, the ratios obtained towards high-mass sources are an order of magnitude lower than towards low-mass sources. Observations of the D/H ratio in CH$_2$DOH around high-mass protostars that have been analyzed in \cite{vanGelder22} and compared to the modeling work of, e.g., \cite{Bögelund18} and \cite{Taquet19}, show that dust temperatures at the prestellar phase of $\geq$~20~K and core ages corresponding to the free-fall time, t$_{\rm ff}$, can explain the observed D/H ratio of CH$_2$DOH in high-mass protostars. In contrast, the models show that dust temperatures of $\leq$~15~K and prestellar core lifetimes of $\geq$~1~t$_{\rm ff}$ are necessary to reproduce the D/H ratios of CH$_2$DOH around low-mass protostars. So a difference in dust temperatures at the prestellar stage can explain the deuterium fractionation pattern in CH$_2$DOH that we observe around protostars. Gas densities in the range of 10$^4$--10$^6$~cm$^{-3}$ were considered in the calculations, and for all t$_{\rm ff}$, high gas densities decrease the D/H ratios more strongly than low gas densities, and more strongly for high dust temperatures. The aforementioned models include the abstraction scheme that deuterates the methyl group of methanol and demonstrate that abstraction at the cold prestellar stage is efficient for deuterating methanol, but its effect is decreasing at warmer dust temperatures.\\
\indent As for CH$_2$DOH, the average of the D/H ratios of CH$_3$OD is strikingly similar for low-mass prestellar cores and low-mass protostars (right panel Fig. \ref{fig:DHratios}). The ratio is clustered around 3\%, which corresponds to the values derived in this work, the lower extreme stems from the upper limit derived towards the prestellar core L1544 (\citealt{Bizzocchi14}). For both, low-mass prestellar cores and low-mass protostars, the average D/H ratio from CH$_3$OD is lower than for CH$_2$DOH. However, the spread towards low-mass protostars is smaller for CH$_3$OD. \\
\indent The average D/H ratios for CH$_2$DOH and CH$_3$OD are strikingly similar for high-mass protostars. However, the spread for CH$_3$OD is smaller; and values do not fall below $\sim$~7~$\times$~10$^{-4}$. This could hint that a mechanism, potentially the aforementioned H-D exchange between methanol and water, might indeed be selectively increasing the abundance of CH$_3$OD around high-mass protostars and altering the value from the prestellar stage. To assess if and to which extent the D/H ratio of CH$_3$OD is altered during the protostellar stage, measurements of the ratio in high-mass prestellar cores are required. However, to date, CH$_3$OD has not been detected in sources of this type.\\
\indent Another caveat in this discussion is the past lack of availability of the spectroscopy and partition function of CH$_3$OD. Its precise spectroscopy plus its partition function have only been published recently (\citealt{Ilyushin24}); and column density derivations of CH$_3$OD have not been conducted uniformly in the past. One option is to derive an estimate for the partition function from the \textbf{asymmetric} rotor approximation (\citealt{Parise04}). Another commonly used approach is to utilize the partition function of CH$_3 ^{18}$OH as a proxy (e.g., \citealt{Jorgensen18}). The latter approach leads to a higher column density of CH$_3$OD. The re-analysis of CH$_3$OD towards IRAS 16293-2422 B has shown that CH$_3$OD is 5.5 times less abundant than previously assumed with the newly published spectroscopy (\citealt{Ilyushin24}).
\begin{figure}
    \centering
    \includegraphics[width=0.99\columnwidth]{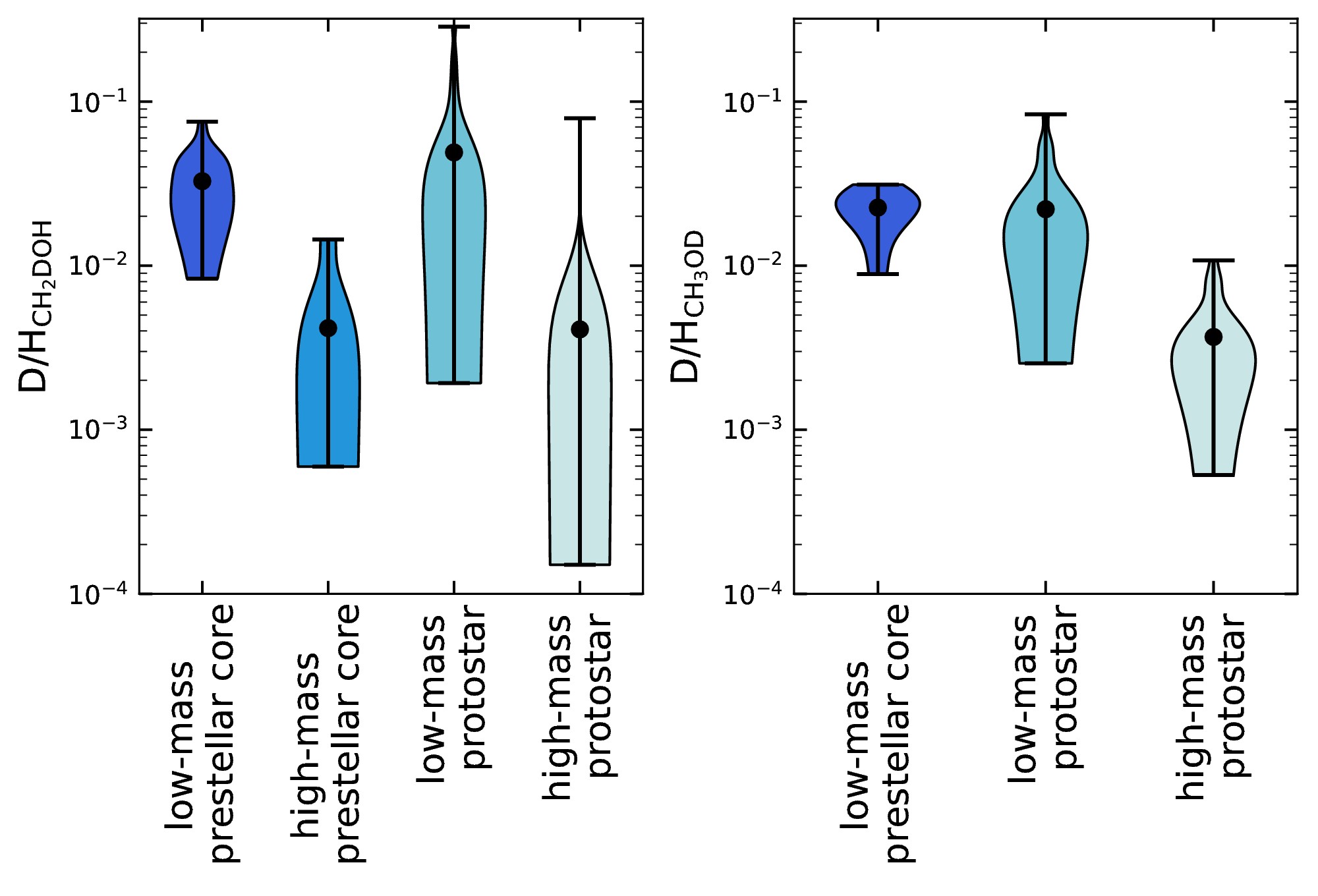}
    \caption{D/H ratios for CH$_2$DOH (left panel) and CH$_3$OD (right panel) across the star-forming sequence. The spread per source type was calculated using all column densities listed in Table \ref{tab:DHreferences}. The black circle denotes the average of the values. The values of CH$_2$DOH have been statistically corrected. Violin plots demonstrate the density distribution of the data: the wider the area of the violin, the more data points are around this value; narrow regions show that data are sparse at this value.}
    \label{fig:DHratios}
\end{figure}

\subsection{CH$_2$DOH/CH$_3$OD Ratio}
Following statistical weighting, the CH$_2$DOH/CH$_3$OD ratio is expected to equal 3. However, compiling this ratio from available data found in the literature shows that it generally deviates from the expected value of 3 (Fig. \ref{fig:CH2DOH_CH3OD}). Low-mass protostars predominantly exceed this value; in the most extreme cases, the ratio is $\sim$~30. On the other hand, the value approaches unity and below for high-mass protostars. In the past, this has been attributed to higher temperatures at the time of formation ($\sim$~30~K) for high-mass protostars (\citealt{Bögelund18}), a detection of CH$_3$OD in a high-mass prestellar core is required to confirm that this is indeed the case. \\
\indent The formation pathways via abstraction and H-D exchange that have been introduced in previous sections are active at different evolutionary stages. The abstraction scheme takes place at the prestellar stage, while H-D exchange requires the presence of a central heating source. Therefore, it is crucial to constrain this ratio at the prestellar stage, which allows us to assess if the ratio of CH$_2$DOH/CH$_3$OD is set at and subsequently inherited from the prestellar stage, or if (and to which degree) it is altered at the protostellar stage. For low-mass sources, the derived ratios agree with the inheritance scenario. The range of CH$_2$DOH/CH$_3$OD ratios is 2.8--8.5 from L1448 and B213-C6 in this work, which agrees with what is seen for most low-mass protostars in the literature, which is 2--20 for most sources in Fig. \ref{fig:CH2DOH_CH3OD}. The ratio for the prestellar core L1544 is $>$ 10 towards its dust peak and $>$3 towards its methanol peak, which is still in agreement with the more extreme values that have been derived for the ratio of CH$_2$DOH/CH$_3$OD for low-mass protostars. In accordance with chemical models (e.g., \citealt{Taquet12,Kulterer22}), the abstraction scheme can account for the range of ratios that are set at the prestellar stage and seen around low-mass protostars. Our data so far suggest that the abstraction path is sufficient to explain the majority of the prestellar core observations, and that the ratio is set at and inherited from the prestellar stage. However, if new data would show a spread in cold cores that deviates from the protostars, we might need to explore the full range of possibilities with the new proposed chemical pathways in the ice. \\
\indent Some of the CH$_3$OD column densities have been caluclated using the partition function of CH$_3 ^{18}$OH, such as in the case of IRAS 16293-2422 B (\citealt{Jorgensen18}). The re-analysis with the newly published spectroscopy data from \cite{Ilyushin24} has led to a decrease in the column density of CH$_3$OD compared to the old value, leading to an increase in the ratio of CH$_2$DOH/CH$_3$OD. This still suggests that CH$_3$OD formation is signifiantly less efficient than that of CH$_2$DOH, and the ratio of CH$_2$DOH/CH$_3$OD is elevated beyond the statistically expected around low-mass sources. How much the newly calculated partition function affects all tabulated values of CH$_3$OD that have been gathered from the literature (Table \ref{tab:DHreferences}) is unclear, but a full re-analysis is beyond the scope of this work. \\
\indent We did compare how much the column density changes if the new partition function values for CH$_3$OD are used for the analysis of our data. For this, we took the reported T$_{\rm ex}$ of the CH$_3$OD data, and then calculated the partition function values based on the new entry in CDMS and the approach used in the original publications, which was either the asymmetric rotor assumption or the use of CH$_3 ^{18}$OH values. The ratio between those two values was then applied to the reported CH$_3$OD column densities. For the majority of the values, the column density did not change by more than 20\%; only in the most extreme cases, it decreased by a factor of 3 for some low-mass protostars. However, this did not change the picture of low-mass sources showing a ratio of CH$_2$DOH/CH$_3$OD that is elevated beyond the statistically expected average of 3 and high-mass sources being below this value.

\begin{figure}
    \centering
    \includegraphics[width=0.99\columnwidth]{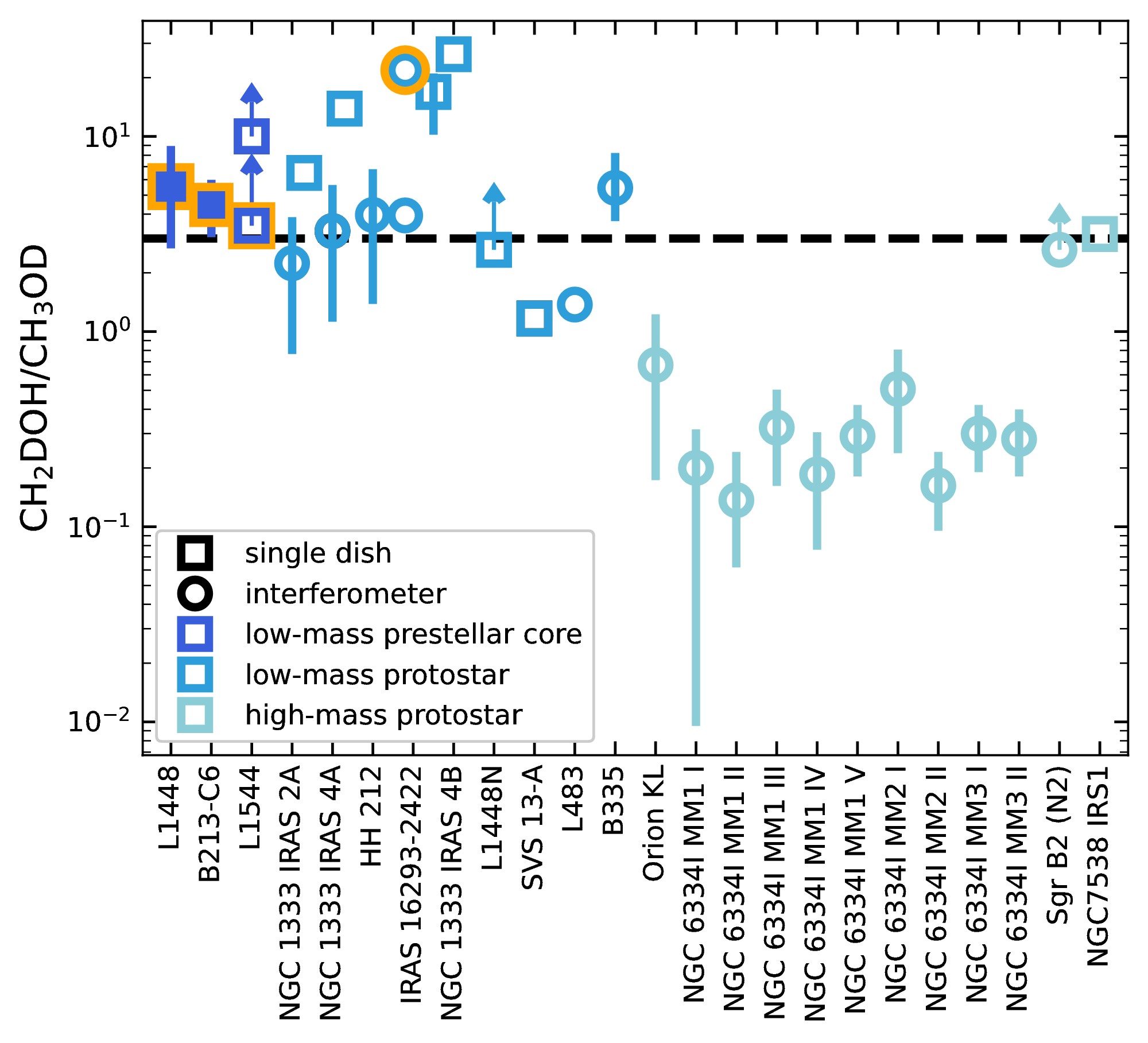}
    \caption{CH$_2$DOH/CH$_3$OD ratio across the star-forming sequence. Dark blue data points represent prestellar cores, sky blue data points depict low-mass protostars, and high-mass protostars are shown in light blue. Squared markers indicate observations with single-dish facilities, circles highlight observations taken with an interferometer. The filled data points show the newly presented values from this work. The column densities used to calculate the remaining ratios and their corresponding references can be found in Table \ref{tab:DHreferences} in the Appendix. The dashed, black line corresponds to the statistically expected value of 3. If multiple data points were available for a source, the range is represented by a line; and the marker shows the average; the lines show the range of values including errors. If a value stems from CH$_3$OD column densities that have been calculated based on the spectroscopy published in \cite{Ilyushin24}, it is framed in orange.}
    \label{fig:CH2DOH_CH3OD}
\end{figure}

\section{Conclusions}\label{sect:conclusion}
We have presented the first detection of CH$_3$OD at the prestellar stage obtained with the IRAM 30m telescope. This allowed us to discuss the isotopic ratios of both singly deuterated methanol isotopologues from prestellar cores to protostars for the low-mass and the high-mass case. We have also discussed how different formation and destruction pathways can alter the isotopic ratios. Moreover, by comparing our obtained D/H ratios of the singly deuterated methanol isotopologues and the CH$_2$DOH/CH$_3$OD ratio with findings from the literature, we have speculated which formation pathways could be dominating during the star formation process.
\begin{enumerate}
    \item For the first time, we present values for the ratio of CH$_2$DOH/CH$_3$OD at the prestellar stage. Depending on the assumed excitation temperature (6--10~K), the ratio is between 2.8 and 8.5 in L1448 in the Perseus star-forming region, and $\leq$ 5.7 in B213-C6 in Taurus; however, the latter is based on a tentative detection. These values fall within the range of ratios that are observed in low-mass protostars. 
    \item The D/H ratios of CH$_3$OD for the two low-mass prestellar cores in this work are similar. The ratio is between 1.4\% and 4.4\% in L1448, and $\leq$ 3.8\% in B213-C6. The D/H ratio of CH$_2$DOH is slightly higher in L1448. It varies between 3.6\% and 6.8\% for L1448, and between 2.4\% and 5.8\% for B213-C6. For both molecules and sources, they are in line with what is found around low-mass protostars. This strongly suggests that lower temperatures at the prestellar stage translate to deuteration levels found in species formed in prestellar ices such as methanol. In accordance with models, the D/H ratios are lower when temperatures are warmer ($>$ 15~K) at the prestellar stage. 
    \item Currently available data hints that the isotopic ratios of deuterated methanol are predominately set at the prestellar stage for the low-mass star-formation case; at least in the case of CH$_2$DOH, the evidence is firm. The deuteration observed in high-mass protostars differs compared to their low-mass counterparts, but if this is due to a lower deuteration efficiency at the prestellar stage or due to other chemical reactions taking place once a central heating source is present can only be determined if CH$_3$OD is also investigated in high-mass prestellar cores. To date, CH$_3$OD has not been detected in high-mass prestellar cores.
    \item So far, CH$_3$OD has only firmly been detected towards one low-mass prestellar core and tentatively towards a second prestellar core. To put a firm constraint on whether the isotopic ratios of methanol are set at and inherited from the prestellar stage, this sample size has to be increased to allow a meaningful statistical comparison, even though current evidence firmly points into that direction.
    \item Due to deuteration being most efficient in the cold, prestellar stage, where the bulk of methanol ice forms, it is expected that the bulk of deuterated methanol is formed at and subsequently inherited from prestellar stages. Selective deuteration of the CH$_3$-group of methanol can explain the observed deuterium fractionation patterns. Chemical abstraction reactions match the here presented data best. However, if new data reveals outliers in the isotopic ratios, it might be necessary to investigate formation pathways that are yet unexplored, such as methanol formation from methane ice or via O insertion into methane. 
    \item All discussed abundances in this work have been derived from observations in the gas phase, while the majority of the discussed formation pathways take place in the ice. It is generally assumed that the isotopic ratios in the ice are equal to the isotopic ratios that are observed in the gas. To verify this, observations of solid deuterated methanol are required. The infrared spectrum of CH$_2$DOH is available (\citealt{Scire19}), but not of CH$_3$OD. However, the detection of CH$_2$DOH in the ices in prestellar cores or protostellar envelopes would push the capability of today's instruments and would be difficult to achieve even with the \textit{James Webb Space Telescope}.
\end{enumerate}

\begin{acknowledgements}
    We thank the anonymous referee for their feedback, which
significantly improved the manuscript. This work is based on observations carried out as part of the large program Gas phase Elemental abundances in Molecular CloudS (GEMS) and under project number 071-23 with the IRAM 30m telescope. IRAM is supported by INSU/CNRS (France), MPG (Germany) and IGN (Spain). We sincerely thank the operators of the IRAM 30m telescope for their assistance with the observations. B.M.K acknowledges the SNSF Postdoc.Mobility stipend P500PT\_214459. This project is co-funded by the European Union (ERC, SUL4LIFE, grant agreement No 101096293). AF, GE, and MRB also thank project PID2022-137980NB-I00 funded by the Spanish
Ministry of Science and Innovation/State Agency of Research MCIN/AEI/10.13039/501100011033 and by “ERDF A way of making Europe”. M.N.D. acknowledges the Holcim Foundation Stipend.
\end{acknowledgements}

\vspace{5mm}
\facilities{IRAM 30m}

\software{\textsc{cassis} (\citealt{Vastel15}), \textsc{class}(\url{ http://www.iram.fr/IRAMFR/GILDAS}), \\ \textsc{gildas-class pipeline} (\citealt{Megias23}), \textsc{matplotlib} (\citealt{Hunter07}), \textsc{numpy} (\citealt{vanderWalt11}), \textsc{scipy} (\citealt{scipy})}

\appendix
\section{Additional Material}
\setcounter{table}{0}
\renewcommand{\thetable}{A\arabic{table}}

In this section, we present the data that we compiled from the literature to create the plot of the D/H ratios in CH$_2$DOH and CH$_3$OD across the star-forming sequence (Fig. \ref{fig:DHratios}) and the plot highlighting available ratios of CH$_2$DOH/CH$_3$OD (Fig. \ref{fig:CH2DOH_CH3OD}). All utilized values with their references and notes on how and at which position in the analyzed systems the data were taken are in Table \ref{tab:DHreferences}. Table \ref{tab:DHreferences} has been adapted from \cite{Drozdovskaya21} and updated with all detections of CH$_2$DOH and CH$_3$OD since its publication. We also list the partition function values of CH$_2$DOH and CH$_3$OD that we used to calculate the column densities in Table \ref{tab:pfval}. In addition, we show the HNC transition at 90.6636~GHz in L1448 and B213-C6, its feature due to the folding of the data in the automated IRAM 30m data reduction pipeline, and the CH$_3$OD transition that might have been affected in Figs. \ref{fig:freqthrow} and \ref{fig:freqthrow2}.

\begin{longrotatetable}
\begin{deluxetable*}{llccclll}
\tablecaption{List of the column densities that were used to calculate the D/H ratios and the ratio of CH$_2$DOH/CH$_3$OD in Figs. \ref{fig:DHratios} and \ref{fig:CH2DOH_CH3OD}. In the column ``Dish type'', SD stands for single dish, and IFM indicates that an interferometer was used to carry out the observations. If $^{13}$CH$_3$OH or CH$_3 ^{18}$OH is mentioned in the ``Note'' column, it means that the column density of the main methanol isotopologue was derived from the specified minor isotopologue. ``Core'' always refers to a ``prestellar core''. The reader is referred to the individual publications for the applied isotopic ratios of $^{12}$C/$^{13}$C and $^{16}$O/$^{18}$O. ``PF P04'' indicates that the partition function approach in \cite{Parise04} was used to determine the column density of CH$_3$OD, ``PF J18'' indicates that approach from \cite{Jorgensen18} was used. The references according to the numbers listed in the column ``Refs'' are the following: (1) \cite{ChaconTanarro19}, (2) \cite{Bizzocchi14}, (3) \cite{Lattanzi20}, (4) \cite{Lin23}, (5) \cite{Ambrose21}, (6) \cite{Fontani15}, (7) \cite{Codella12}, (8) \cite{Taquet19}, (9) \cite{Parise06}, (10) \cite{Sahu19}, (11) \cite{Bianchi17HH212}, (12) \cite{Lee19}, (13) \cite{Parise02}, (14) \cite{Manigand20}, (15) \cite{Jorgensen18}, (16) \cite{Ilyushin24}, (17) \cite{Bianchi17SVS13}, (18) \cite{Agundez19}, (19) \cite{Hsu20}, (20) \cite{vanGelder20}, (21) \cite{OspinaZamudio18}, (22) \cite{Ligterink21}, (23) \cite{Okoda23}, (24) \cite{Imai22}, (25) \cite{Mercimek22}, (26) \cite{Chahine22}, (27) \cite{MartinDomenech21} (28) \cite{Lee23}, (29) \cite{Okoda24}, (30) \cite{Neill13}, (31) \cite{Peng12}, (32) \cite{Bögelund18}, (33) \cite{Dartois00}, (34) \cite{Belloche16}, (35) \cite{OspinaZamudio19}, (36) \cite{vanderWalt23}, (37) \cite{vanGelder22}, (*) private communication B. Kulterer. Errors on the column densities are generally not larger than 20--25\%; for the specific values the reader is referred to the original publications.}\label{tab:DHreferences}
\tablehead{
\colhead{Name} & \colhead{Type} & 
\colhead{N(CH$_3$OH)} & \colhead{N(CH$_2$DOH)} & 
\colhead{N(CH$_3$OD)} & \colhead{Dish type} & 
\colhead{Note} & \colhead{Ref.}  \\
\colhead{} & \colhead{} & \colhead{(cm$^{-2}$)} & \colhead{(cm$^{-2}$)} & \colhead{(cm$^{-2}$)} & \colhead{} & \colhead{} & \colhead{} \\}
\startdata
    L1448 & low-mass core & 8.18~$\times$~10$^{13}$ & 1.43~$\times$~10$^{13}$ & 2.89~$\times$~10$^{12}$ & SD & T$_{\rm ex}$ = 6~K & this work \\
    L1448 & low-mass core & 8.18~$\times$~10$^{13}$ & 9.47 $\times$ 10$^{12}$ &  2.54~$\times$ 10$^{12}$ & SD & T$_{\rm ex}$ = 8 K & this work \\ 
    L1448 & low-mass core & 8.18~$\times$~10$^{13}$ & 7.96 $\times$ 10$^{12}$ &  2.54~$\times$ 10$^{12}$ & SD & T$_{\rm ex}$ = 10 K & this work \\ 
    B213-C6 & low-mass core & 3.50~$\times$~10$^{13}$ & 7.91 $\times$ 10$^{12}$ &  1.06~$\times$ 10$^{12}$ & SD & T$_{\rm ex}$ = 6 K & this work \\ 
    B213-C6 & low-mass core & 3.50~$\times$~10$^{13}$ & 5.28 $\times$ 10$^{12}$ &  1.04~$\times$ 10$^{12}$ & SD & T$_{\rm ex}$ = 8 K & this work \\ 
    B213-C6 & low-mass core & 3.50~$\times$~10$^{13}$ & 4.55 $\times$ 10$^{12}$ &  1.10~$\times$ 10$^{12}$ & SD & T$_{\rm ex}$ = 10 K & this work \\    
    L1544  & low-mass core & 3.90~$\times$~10$^{13}$  & 2.80~$\times$~10$^{12}$ & - & SD & dust peak &  (1) \\
    L1544  & low-mass core & 5.90~$\times$~10$^{13}$  & 3.30~$\times$~10$^{12}$ & - & SD & CH$_3$OH peak &  (1),(*) \\
    L1544  & low-mass core & 2.70~$\times$~10$^{13}$  & 2.40~$\times$~10$^{12}$ & $<$ 2.40 $\times$ 10$^{11}$ & SD & dust peak &  (2) \\
    L183  & low-mass core & 4.90~$\times$~10$^{13}$  & 1.90~$\times$~10$^{12}$ & - & SD &  &  (3) \\
    H-MM1  & low-mass core & 3.20~$\times$~10$^{13}$  & 1.80~$\times$~10$^{12}$ & - & SD &  &  (4) \\
    L694-2  & low-mass core & 6.00~$\times$~10$^{13}$  & 1.50~$\times$~10$^{12}$ & - & SD &  & (4) \\
    L1495-B10 Seo06  & low-mass core & 2.60~$\times$~10$^{13}$  & 1.00~$\times$~10$^{12}$ & - & SD &  &  (5) \\
    L1495-B10 Seo07  & low-mass core & 1.00~$\times$~10$^{13}$  & 1.35~$\times$~10$^{12}$ & - & SD &  &  (5) \\
    L1495-B10 Seo08  & low-mass core & 1.40~$\times$~10$^{13}$  & 1.98~$\times$~10$^{12}$ & - & SD &  &  (5) \\
    L1495-B10 Seo09  & low-mass core & 2.30~$\times$~10$^{13}$  & 2.87~$\times$~10$^{12}$ & - & SD &  &  (5) \\
    L1495-B10 Seo10  & low-mass core & 2.20~$\times$~10$^{13}$  & 2.49~$\times$~10$^{12}$ & - & SD &  &  (5) \\
    L1495-B10 Seo12  & low-mass core & 2.90~$\times$~10$^{13}$  & 1.63~$\times$~10$^{12}$ & - & SD &  &  (5) \\
    L1495-B10 Seo15  & low-mass core & 1.50~$\times$~10$^{13}$  & 1.57~$\times$~10$^{12}$ & - & SD &  &  (5) \\
    L1495-B10 Seo16  & low-mass core & 1.60~$\times$~10$^{13}$  & 1.26~$\times$~10$^{12}$ & - & SD &  &  (5) \\
    L1495-B10 Seo06  & low-mass core & 8.40~$\times$~10$^{12}$  & 1.90~$\times$~10$^{12}$ & - & SD &  &  (5) \\
    AFGL5142-EC & high-mass core & 6.15~$\times$~10$^{15}$  & 1.10~$\times$~10$^{13}$ & - & SD &  &  (6) \\
    AFGL5142-EC & high-mass core & 4.06~$\times$~10$^{15}$  & 1.10~$\times$~10$^{13}$ & - & SD & $^{13}$CH$_3$OH &  (6) \\
    05358-mm3 & high-mass core & 2.49~$\times$~10$^{15}$  & 8.00~$\times$~10$^{12}$ & - & SD &  &  (6) \\
    05358-mm3 & high-mass core & 1.18~$\times$~10$^{15}$  & 8.00~$\times$~10$^{12}$ & - & SD & $^{13}$CH$_3$OH &  (6) \\
    G034-G2 (MM2)  & high-mass core & 1.75~$\times$~10$^{14}$  & 3.00~$\times$~10$^{12}$ & - & SD &  &  (6) \\
    G034-G2 (MM2) & high-mass core & 6.93~$\times$~10$^{13}$  & 3.00~$\times$~10$^{12}$ & - & SD & $^{13}$CH$_3$OH &  (6) \\
    L1157-B1 & low-mass protostar & 2.31~$\times$~10$^{15}$  & 4.00~$\times$~10$^{13}$ & - & SD & C$^{13}$CH$_3$OH, shocked region &  (7) \\
    NGC 1333 IRAS2A & low-mass protostar & 5.00~$\times$~10$^{18}$  & 2.90~$\times$~10$^{17}$ & 7.90 $\times$ 10$^{16}$ & IFM & PF P04 &  (8) \\   
    NGC 1333 IRAS2A & low-mass protostar & 5.00~$\times$~10$^{18}$  & 2.90~$\times$~10$^{17}$ & 3.60 $\times$ 10$^{16}$ & IFM & PF J18 &  (8) \\ 
    NGC 1333 IRAS2A & low-mass protostar & 1.01~$\times$~10$^{15}$  & 5.20~$\times$~10$^{14}$ & $<$ 8.00 $\times$ 10$^{13}$ & SD &  &  (9) \\   
    NGC 1333 IRAS4A & low-mass protostar & 1.60~$\times$~10$^{19}$  & 5.90~$\times$~10$^{17}$ & 1.10 $\times$ 10$^{17}$ & IFM & PF P04 &  (8) \\   
    NGC 1333 IRAS4A & low-mass protostar & 1.60~$\times$~10$^{19}$  & 5.90~$\times$~10$^{17}$ & 5.00 $\times$ 10$^{17}$ & IFM & PF J18 &  (8) \\ 
    NGC 1333 IRAS4A & low-mass protostar & 6.90~$\times$~10$^{14}$  & 4.30~$\times$~10$^{14}$ & 3.10 $\times$ 10$^{13}$ & SD &  &  (9) \\   
    NGC 1333 IRAS4A1 & low-mass protostar & 1.34~$\times$~10$^{17}$  & 6.51~$\times$~10$^{15}$ & -  & IFM & $^{13}$CH$_3$OH &  (10) \\ 
    NGC 1333 IRAS4A2 & low-mass protostar & 2.25~$\times$~10$^{19}$  & 1.30~$\times$~10$^{17}$ & - & IFM & $^{13}$CH$_3$OH &  (10) \\ 
    HH 212 & low-mass protostar & 2.24~$\times$~10$^{18}$  & 6.40~$\times$~10$^{16}$ & 9.90 $\times$ 10$^{15}$ & IFM & PF P04,$^{13}$CH$_3$OH &  (8) \\   
    HH 212 & low-mass protostar & 1.60~$\times$~10$^{19}$  & 5.90~$\times$~10$^{17}$ & 5.00 $\times$ 10$^{17}$ & IFM & PF J18, $^{13}$CH$_3$OH &  (8) \\ 
    HH 212 & low-mass protostar & 4.55~$\times$~10$^{18}$  & 1.10~$\times$~10$^{17}$ & - & IFM & $^{13}$CH$_3$OH & (11) \\
    HH 212 & low-mass protostar & 1.25~$\times$~10$^{18}$  & 1.60~$\times$~10$^{17}$ & - & IFM & $^{13}$CH$_3$OH, lower disc atmosphere & (12) \\  
    HH 212 & low-mass protostar & 3.40~$\times$~10$^{17}$  & 9.20~$\times$~10$^{16}$ & - & IFM & $^{13}$CH$_3$OH, lower disc atmosphere & (12) \\     
    IRAS 16293-2422 & low-mass protostar & 3.50~$\times$~10$^{15}$  & 3.00~$\times$~10$^{15}$ & 1.50 $\times$ 10$^{14}$ & SD & 10$^{\prime\prime}$ circumbinary envelope, 20K & (13) \\
    IRAS 16293-2422 & low-mass protostar & 3.50~$\times$~10$^{15}$  & 3.00~$\times$~10$^{15}$ & 2.80 $\times$ 10$^{14}$ & SD & 10$^{\prime\prime}$ circumbinary envelope, 48K & (13) \\    
    IRAS 16293-2422 & low-mass protostar & 9.80~$\times$~10$^{15}$  & 3.00~$\times$~10$^{15}$ & 1.50 $\times$ 10$^{14}$ & SD & 10$^{\prime\prime}$ circumbinary envelope, $^{13}$CH$_3$OH & (13) \\
    IRAS 16293-2422 & low-mass protostar & 1.30~$\times$~10$^{19}$  & 2.80~$\times$~10$^{17}$ & 2.80 $\times$ 10$^{17}$ & IFM & 0.6$^{\prime\prime}$ offset NE from A & (14) \\
    IRAS 16293-2422 & low-mass protostar & 1.36~$\times$~10$^{19}$  & 1.10~$\times$~10$^{18}$ & 2.80 $\times$ 10$^{17}$ & IFM & 0.6$^{\prime\prime}$ offset NE from A, $^{13}$CH$_3$OH & (14) \\
    IRAS 16293-2422 & low-mass protostar & 1.28~$\times$~10$^{19}$  & 1.10~$\times$~10$^{17}$ & 2.80 $\times$ 10$^{17}$ & IFM & 0.6$^{\prime\prime}$ offset NE from A, CH$_3 ^{18}$OH & (14) \\
    IRAS 16293-2422 & low-mass protostar & 1.00~$\times$~10$^{19}$  & 7.10~$\times$~10$^{17}$ & 1.80 $\times$ 10$^{17}$ & IFM & 0.5$^{\prime\prime}$ offset SW from B, CH$_3 ^{18}$OH & (15) \\
    IRAS 16293-2422 & low-mass protostar & 1.28~$\times$~10$^{19}$  & 7.10~$\times$~10$^{17}$ & 3.25 $\times$ 10$^{16}$ & IFM & CH$_3 ^{18}$OH & (16) \\
    NGC 1333 IRAS4B & low-mass protostar & 8.00~$\times$~10$^{14}$  & 2.90~$\times$~10$^{14}$ & 1.10 $\times$ 10$^{13}$ & SD & 20$^{\prime\prime}$ circumbinary envelope & (9) \\
    SVS13-A & low-mass protostar & 1.09~$\times$~10$^{17}$  & 7.00~$\times$~10$^{14}$ & 6.00 $\times$ 10$^{14}$ & SD & low-T 3$^{\prime\prime}$ component, $^{13}$CH$_3$OH & (17) \\
    SVS13-A & low-mass protostar & 1.39~$\times$~10$^{18}$  & 4.00~$\times$~10$^{17}$ & 6.00 $\times$ 10$^{14}$ & SD & high-T 0.3$^{\prime\prime}$ component, $^{13}$CH$_3$OH & (17) \\
    L483 & low-mass protostar & 2.92~$\times$~10$^{14}$  & 5.50~$\times$~10$^{12}$ & 4.00 $\times$ 10$^{12}$ & IFM & $^{13}$CH$_3$OH & (18) \\
    G211.47-19.27S & low-mass protostar & 6.44~$\times$~10$^{17}$  & 2.30~$\times$~10$^{16}$ & - & IFM & $^{13}$CH$_3$OH & (19) \\   
    B1-c & low-mass protostar & 1.80~$\times$~10$^{18}$  & 1.03~$\times$~10$^{17}$ & - & IFM & $^{13}$CH$_3$OH, ALMA Band 3 & (20) \\ 
    B1-c & low-mass protostar & 1.90~$\times$~10$^{18}$  & 1.60~$\times$~10$^{17}$ & - & IFM & CH$_3 ^{18}$OH, ALMA Band 6 & (20) \\ 
    B1-c & low-mass protostar & 1.26~$\times$~10$^{18}$  & 1.60~$\times$~10$^{17}$ & - & IFM & $^{13}$CH$_3$OH, warm component & (20) \\ 
    B1-c & low-mass protostar & 1.90~$\times$~10$^{18}$  & 1.60~$\times$~10$^{17}$ & - & IFM & CH$_3 ^{18}$OH, warm component & (20) \\ 
    S68N & low-mass protostar & 1.40~$\times$~10$^{18}$  & 6.02~$\times$~10$^{16}$ & - & IFM & CH$_3 ^{18}$OH, ALMA Band 6 & (20) \\ 
    S68N & low-mass protostar & 7.00~$\times$~10$^{17}$  & 6.00~$\times$~10$^{16}$ & - & IFM & $^{13}$CH$_3$OH, warm component & (20) \\ 
    S68N & low-mass protostar & 1.40~$\times$~10$^{18}$  & 6.00~$\times$~10$^{16}$ & - & IFM & CH$_3 ^{18}$OH, warm component & (20) \\ 
    CepE-mm & low-mass protostar & 4.90~$\times$~10$^{17}$  & 1.77~$\times$~10$^{16}$ & - & IFM & $^{13}$CH$_3$OH & (21) \\ 
    SMM1-a & low-mass protostar & 7.84~$\times$~10$^{17}$  & 2.30~$\times$~10$^{16}$ & - & IFM & CH$_3 ^{18}$OH & (22) \\ 
    SMM1-a & low-mass protostar & 5.30~$\times$~10$^{17}$  & 2.30~$\times$~10$^{16}$ & - & IFM & $^{13}$CH$_3$OH & (22) \\ 
    IRAS15398-3359 & low-mass protostar & 3.20~$\times$~10$^{18}$  & 3.05~$\times$~10$^{17}$ & - & IFM &  & (23) \\ 
    IRAS16544-1604 & low-mass protostar & 2.70~$\times$~10$^{18}$  & 1.20~$\times$~10$^{17}$ & - & IFM &  & (24) \\ 
    B5-IRS1 & low-mass protostar & 1.50~$\times$~10$^{13}$  & 5.50~$\times$~10$^{12}$ & - & SD & envelope & (25) \\ 
    L1455-IRS1 & low-mass protostar & 7.50~$\times$~10$^{13}$  & 4.00~$\times$~10$^{12}$ & - & SD & hot corino & (25) \\ 
    L1551-IRS5 & low-mass protostar & 1.60~$\times$~10$^{14}$  & 2.00~$\times$~10$^{13}$ & - & SD & envelope & (25) \\    
    HOPS-108 & low-mass protostar & 9.50~$\times$~10$^{18}$  & 2.00~$\times$~10$^{17}$ & - & IFM &  & (26) \\ 
    Ser-emb-11W & low-mass protostar & 3.70~$\times$~10$^{18}$  & 4.40~$\times$~10$^{17}$ & - & IFM &  & (27) \\ 
    HOPS-373SW & low-mass protostar & 8.04~$\times$~10$^{17}$  & 4.03~$\times$~10$^{17}$ & - & IFM &  & (28) \\ 
    B335& low-mass protostar & 4.40~$\times$~10$^{18}$  & 4.70~$\times$~10$^{17}$ & 6.00~$\times$~10$^{16}$ & IFM & 0.1$^{\prime\prime}$ offset & (29) \\ 
    B335& low-mass protostar & 9.00~$\times$~10$^{18}$  & 8.50~$\times$~10$^{17}$ & 2.20~$\times$~10$^{17}$ & IFM & 0.06$^{\prime\prime}$ offset & (29) \\ 
    B335& low-mass protostar & 1.30~$\times$~10$^{19}$  & 2.50~$\times$~10$^{18}$ & 4.50~$\times$~10$^{17}$ & IFM & -0.03$^{\prime\prime}$ offset & (29) \\ 
    B335& low-mass protostar & 8.00~$\times$~10$^{18}$  & 1.00~$\times$~10$^{18}$ & 1.90~$\times$~10$^{17}$ & IFM & -0.06$^{\prime\prime}$ offset & (29) \\ 
    B335& low-mass protostar & 3.70~$\times$~10$^{18}$  & 5.20~$\times$~10$^{17}$ & 1.10~$\times$~10$^{17}$ & IFM & -0.1$^{\prime\prime}$ offset & (29) \\ 
    Orion KL & high-mass protostar & 6.00~$\times$~10$^{17}$  & 3.50~$\times$~10$^{15}$ & 3.00~$\times$~10$^{15}$ & IFM & compact ridge, $^{13}$CH$_3$OH & (30) \\ 
    Orion KL dM1 & high-mass protostar & 2.10~$\times$~10$^{18}$  & 2.40~$\times$~10$^{15}$ & - & IFM & & (31) \\ 
    Orion KL dM2 & high-mass protostar & 1.90~$\times$~10$^{18}$  & 1.40~$\times$~10$^{15}$ & - & IFM & & (31) \\ 
    Orion KL dM3 & high-mass protostar & 1.70~$\times$~10$^{18}$  & 1.50~$\times$~10$^{15}$ & - & IFM & & (31) \\ 
    Orion KL IRc2 & high-mass protostar & 4.70~$\times$~10$^{17}$  & 4.40~$\times$~10$^{15}$ & - & IFM & & (31) \\ 
    NGC 6334I MM1 I & high-mass protostar & 5.15~$\times$~10$^{19}$  & 1.10~$\times$~10$^{17}$ & 5.50 $\times$ 10$^{17}$ & IFM & $^{13}$CH$_3$OH & (32) \\ 
    NGC 6334I MM1 I & high-mass protostar & 1.22~$\times$~10$^{20}$  & 1.10~$\times$~10$^{17}$ & 5.50 $\times$ 10$^{17}$ & IFM & CH$_3 ^{18}$OH & (32) \\ 
    NGC 6334I MM1 II & high-mass protostar & 4.59~$\times$~10$^{19}$  & 5.20~$\times$~10$^{16}$ & 3.80 $\times$ 10$^{17}$ & IFM & $^{13}$CH$_3$OH & (32) \\ 
    NGC 6334I MM1 II & high-mass protostar & 1.08~$\times$~10$^{20}$  & 5.20~$\times$~10$^{16}$ & 3.80 $\times$ 10$^{17}$ & IFM & CH$_3 ^{18}$OH & (32) \\     
    NGC 6334I MM1 III & high-mass protostar & 5.15~$\times$~10$^{19}$  & 7.40~$\times$~10$^{16}$ & 2.30 $\times$ 10$^{17}$ & IFM & $^{13}$CH$_3$OH & (32) \\ 
    NGC 6334I MM1 III & high-mass protostar & 7.65~$\times$~10$^{19}$  & 7.40~$\times$~10$^{16}$ & 2.30 $\times$ 10$^{17}$ & IFM & CH$_3 ^{18}$OH & (32) \\    
    NGC 6334I MM1 IV & high-mass protostar & 3.22~$\times$~10$^{19}$  & 3.15~$\times$~10$^{16}$ & 1.70 $\times$ 10$^{17}$ & IFM & $^{13}$CH$_3$OH & (32) \\ 
    NGC 6334I MM1 IV & high-mass protostar & 6.75~$\times$~10$^{19}$  & 3.15~$\times$~10$^{16}$ & 1.70 $\times$ 10$^{17}$ & IFM & CH$_3 ^{18}$OH & (32) \\    
    NGC 6334I MM1 V & high-mass protostar & 8.06~$\times$~10$^{18}$  & 1.25~$\times$~10$^{16}$ & 4.30 $\times$ 10$^{16}$ & IFM & $^{13}$CH$_3$OH & (32) \\ 
    NGC 6334I MM1 V & high-mass protostar & 1.84~$\times$~10$^{19}$  & 1.25~$\times$~10$^{16}$ & 4.30 $\times$ 10$^{16}$ & IFM & CH$_3 ^{18}$OH & (32) \\    
    NGC 6334I MM2 I & high-mass protostar & 4.09~$\times$~10$^{19}$  & 9.15~$\times$~10$^{16}$ & 1.80 $\times$ 10$^{17}$ & IFM & $^{13}$CH$_3$OH & (32) \\ 
    NGC 6334I MM2 I & high-mass protostar & 5.18~$\times$~10$^{19}$  & 9.15~$\times$~10$^{16}$ & 1.80 $\times$ 10$^{17}$ & IFM & CH$_3 ^{18}$OH & (32) \\ 
    NGC 6334I MM2 II & high-mass protostar & 1.12~$\times$~10$^{19}$  & 6.50~$\times$~10$^{15}$ & 4.00 $\times$ 10$^{16}$ & IFM & $^{13}$CH$_3$OH & (32) \\ 
    NGC 6334I MM2 II & high-mass protostar & 1.44~$\times$~10$^{19}$  & 6.50~$\times$~10$^{15}$ & 4.00 $\times$ 10$^{16}$ & IFM & CH$_3 ^{18}$OH & (32) \\ 
    NGC 6334I MM3 I & high-mass protostar & 5.58~$\times$~10$^{18}$  & 4.50~$\times$~10$^{15}$ & 1.50 $\times$ 10$^{16}$ & IFM & $^{13}$CH$_3$OH & (32) \\ 
    NGC 6334I MM3 I & high-mass protostar & 6.30~$\times$~10$^{18}$  & 4.50~$\times$~10$^{15}$ & 1.50 $\times$ 10$^{16}$ & IFM & CH$_3 ^{18}$OH & (32) \\ 
    NGC 6334I MM3 II & high-mass protostar & 4.96~$\times$~10$^{18}$  & 4.50~$\times$~10$^{15}$ & 1.60 $\times$ 10$^{16}$ & IFM & $^{13}$CH$_3$OH  & (32) \\ 
    NGC 6334I MM3 II & high-mass protostar & 5.85~$\times$~10$^{18}$  & 4.50~$\times$~10$^{15}$ & 1.60 $\times$ 10$^{16}$ & IFM & CH$_3 ^{18}$OH & (32) \\ 
    RAFGL 7009S & high-mass protostar & 1.20~$\times$~10$^{16}$  & - & 9.50 $\times$ 10$^{13}$ & IFM &  & (33) \\ 
    SgrB2 (N2) & high-mass protostar & 4.00~$\times$~10$^{19}$  & 4.80 $\times$ 10$^{16}$ & $<$ 2.60 $\times$ 10$^{16}$ & IFM &  & (34) \\ 
    NGC7538-IRS1 & high-mass protostar & 3.78~$\times$~10$^{17}$  & 1.20~$\times$~10$^{15}$ & 3.80 $\times$ 10$^{14}$ & SD & $^{13}$CH$_3$OH & (35) \\ 
    AFGL5142-MM & high-mass protostar & 2.63~$\times$~10$^{16}$  & 1.06~$\times$~10$^{15}$ & - & SD &  & (6) \\ 
    AFGL5142-MM & high-mass protostar & 4.47~$\times$~10$^{15}$  & 1.06~$\times$~10$^{15}$ & - & SD & $^{13}$CH$_3$OH & (6) \\ 
    18089-1732  & high-mass protostar & 3.18~$\times$~10$^{16}$  & 9.00~$\times$~10$^{14}$ & - & SD &  & (6) \\ 
    18089-1732  & high-mass protostar & 4.94~$\times$~10$^{16}$  & 9.00~$\times$~10$^{14}$ & - & SD & $^{13}$CH$_3$OH & (6) \\
    G75-core  & high-mass protostar & 1.51~$\times$~10$^{16}$  & 5.50~$\times$~10$^{13}$ & - & SD &  & (6) \\ 
    G75-core  & high-mass protostar & 3.23~$\times$~10$^{15}$  & 5.50~$\times$~10$^{13}$ & - & SD & $^{13}$CH$_3$OH & (6) \\
    CygnusX-N63 & high-mass protostar & 5.00~$\times$~10$^{17}$  & 1.50~$\times$~10$^{16}$ & - & SD &  & (36) \\ 
    CygnusX-N63  & high-mass protostar & 3.94~$\times$~10$^{17}$  & 1.50~$\times$~10$^{16}$ & - & SD & $^{13}$CH$_3$OH & (36) \\
    101889 C1 & high-mass protostar & 6.80~$\times$~10$^{17}$  & 8.00~$\times$~10$^{15}$ & - & IFM & $^{13}$CH$_3$OH & (37) \\ 
    615590 C1 & high-mass protostar & 1.40~$\times$~10$^{18}$  & 3.00~$\times$~10$^{16}$ & - & IFM & $^{13}$CH$_3$OH & (37) \\ 
    693050 & high-mass protostar & 1.30~$\times$~10$^{18}$  & 1.10~$\times$~10$^{16}$ & - & IFM & CH$_3 ^{18}$OH & (37) \\ 
    707948 & high-mass protostar & 1.10~$\times$~10$^{19}$  & 2.00~$\times$~10$^{17}$ & - & IFM & CH$_3 ^{18}$OH & (37) \\     
    724566 & high-mass protostar & 4.00~$\times$~10$^{18}$  & 2.00~$\times$~10$^{16}$ & - & IFM & CH$_3 ^{18}$OH  & (37) \\ 
    744757A & high-mass protostar & 1.70~$\times$~10$^{18}$  & 8.00~$\times$~10$^{15}$ & - & IFM & CH$_3 ^{18}$OH & (37) \\ 
    767784 & high-mass protostar & 2.50~$\times$~10$^{18}$  & 1.00~$\times$~10$^{16}$ & - & IFM & CH$_3 ^{18}$OH & (37) \\ 
    800287 & high-mass protostar & 1.80~$\times$~10$^{18}$  & 1.00~$\times$~10$^{16}$ & - & IFM & CH$_3 ^{18}$OH & (37) \\ 
    865468A C1 & high-mass protostar & 1.00~$\times$~10$^{19}$  & 1.50~$\times$~10$^{17}$ & - & IFM & CH$_3 ^{18}$OH & (37) \\ 
    865468A C2 & high-mass protostar & 2.50~$\times$~10$^{18}$  & 6.00~$\times$~10$^{16}$ & - & IFM & $^{13}$CH$_3$OH & (37) \\ 
    865468B & high-mass protostar & 8.60~$\times$~10$^{17}$  & 6.50~$\times$~10$^{15}$ & - & IFM & CH$_3 ^{18}$OH & (37) \\ 
    8811427A & high-mass protostar & 1.10~$\times$~10$^{19}$  & 8.30~$\times$~10$^{16}$ & - & IFM & CH$_3 ^{18}$OH & (37) \\ 
    8811427B & high-mass protostar & 3.40~$\times$~10$^{18}$  & 1.50~$\times$~10$^{16}$ & - & IFM & CH$_3 ^{18}$OH & (37) \\ 
    8811427C & high-mass protostar & 1.10~$\times$~10$^{19}$  & 8.30~$\times$~10$^{16}$ & - & IFM & CH$_3 ^{18}$OH & (37) \\ 
    G023.3891+00.1851 & high-mass protostar & 7.90~$\times$~10$^{17}$  & 9.50~$\times$~10$^{15}$ & - & IFM & CH$_3 ^{18}$OH & (37) \\ 
    G023.6566-00.1273 & high-mass protostar & 7.10~$\times$~10$^{17}$  & 2.50~$\times$~10$^{16}$ & - & IFM & $^{13}$CH$_3$OH  & (37) \\ 
    G025.6498+01.0491 & high-mass protostar & 3.40~$\times$~10$^{18}$  & 1.50~$\times$~10$^{16}$ & - & IFM & CH$_3 ^{18}$OH & (37) \\ 
    G305.2017+00.2072A1 & high-mass protostar & 8.50~$\times$~10$^{17}$  & 5.00~$\times$~10$^{15}$ & - & IFM & CH$_3 ^{18}$OH & (37) \\ 
    G316.6412-00.0867 & high-mass protostar & 3.50~$\times$~10$^{18}$  & 2.80~$\times$~10$^{16}$ & - & IFM & CH$_3 ^{18}$OH & (37) \\ 
    G318.0489+00.0854B & high-mass protostar & 1.90~$\times$~10$^{18}$  & 8.00~$\times$~10$^{15}$ & - & IFM & CH$_3 ^{18}$OH & (37) \\ 
    G318.9480-00.1969A1 & high-mass protostar & 9.60~$\times$~10$^{18}$  & 7.00~$\times$~10$^{16}$ & - & IFM & CH$_3 ^{18}$OH & (37) \\ 
    G323.7399-00.2617B2 & high-mass protostar & 5.70~$\times$~10$^{18}$  & 2.00~$\times$~10$^{16}$ & - & IFM & CH$_3 ^{18}$OH & (37) \\   
    G327.1192+00.5103 & high-mass protostar & 2.90~$\times$~10$^{18}$  & 3.50~$\times$~10$^{16}$ & - & IFM & CH$_3 ^{18}$OH & (37) \\  
    G345.5043+00.3480 C1 & high-mass protostar & 4.00~$\times$~10$^{18}$  & 5.00~$\times$~10$^{16}$ & - & IFM & CH$_3 ^{18}$OH & (37) \\  
    G345.5043+00.3480 C2 & high-mass protostar & 1.70~$\times$~10$^{18}$  & 3.00~$\times$~10$^{16}$ & - & IFM & $^{13}$CH$_3$OH & (37) \\  
\enddata
\end{deluxetable*}
\end{longrotatetable}

\begin{table}[]
    \caption{Partition function values for CH$_2$DOH and CH$_3$OD at 6, 8, and 10 K.}
    \centering
    \begin{tabular}{lll}
   \hline \hline 
    Molecule & T$_{\rm ex}$ (K) & Q(T) \\
    \hline 
    CH$_2$DOH & 6 & 13.19761 \\
    & 8 &  22.44675\\
    & 10 & 34.41106 \\
    CH$_3$OD & 6 & 16.08576 \\
    & 8 & 25.99831 \\
    & 10 & 37.55357 \\
    \hline 
    \end{tabular}
    \label{tab:pfval}
\end{table}

\renewcommand\thefigure{\thesection.\arabic{figure}}    
\setcounter{figure}{0}    
\begin{figure}
    \centering
    \includegraphics[width=0.49\textwidth]{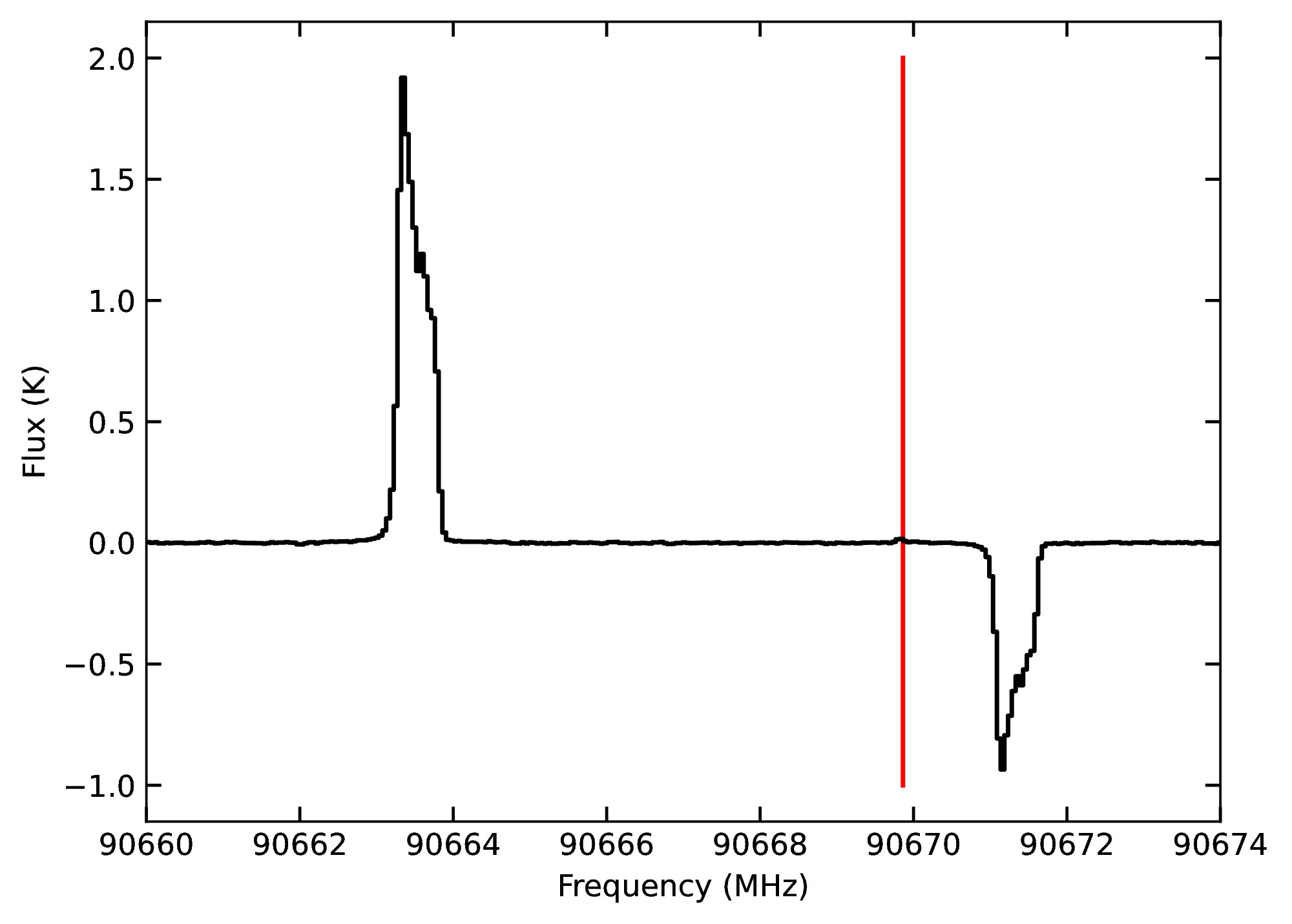}\includegraphics[width=0.49\textwidth]{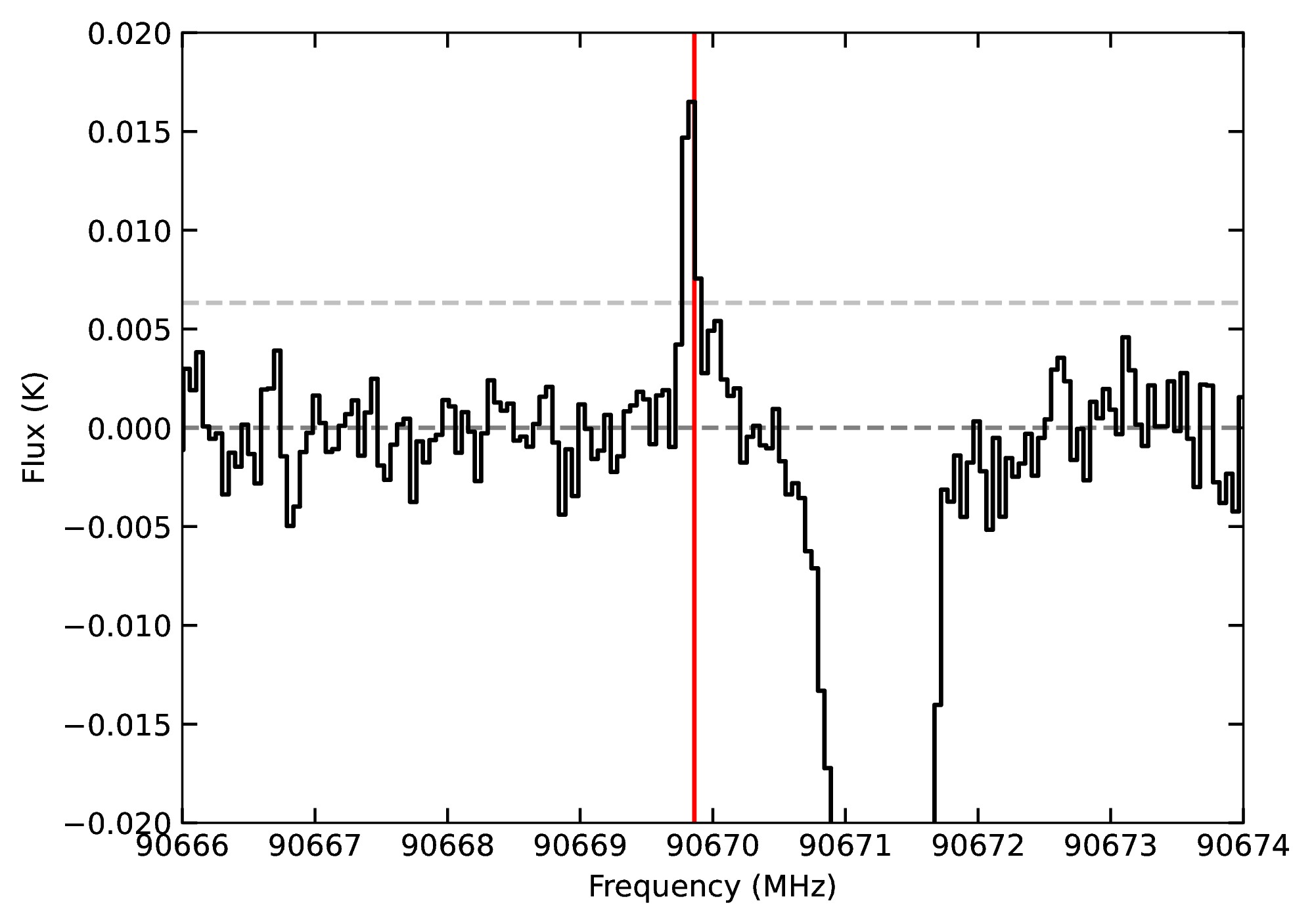}
    \caption{\textit{Left:} HNC transition at 90.6636~GHz and its corresponding absorption feature produced during the folding of the frequency-switched data in L1448. The red line denotes the position of the CH$_3$OD transition at 90.6699~GHz. \textit{Right:} Zoomed-in picture of the image on the left. Again, the red line denotes the CH$_3$OD line that may have been affected; the dashed gray lines represent the flux at 0~K and at 3~$\times$~rms.}
    \label{fig:freqthrow}
\end{figure}

\begin{figure*}
    \centering
    \includegraphics[width=0.49\textwidth]{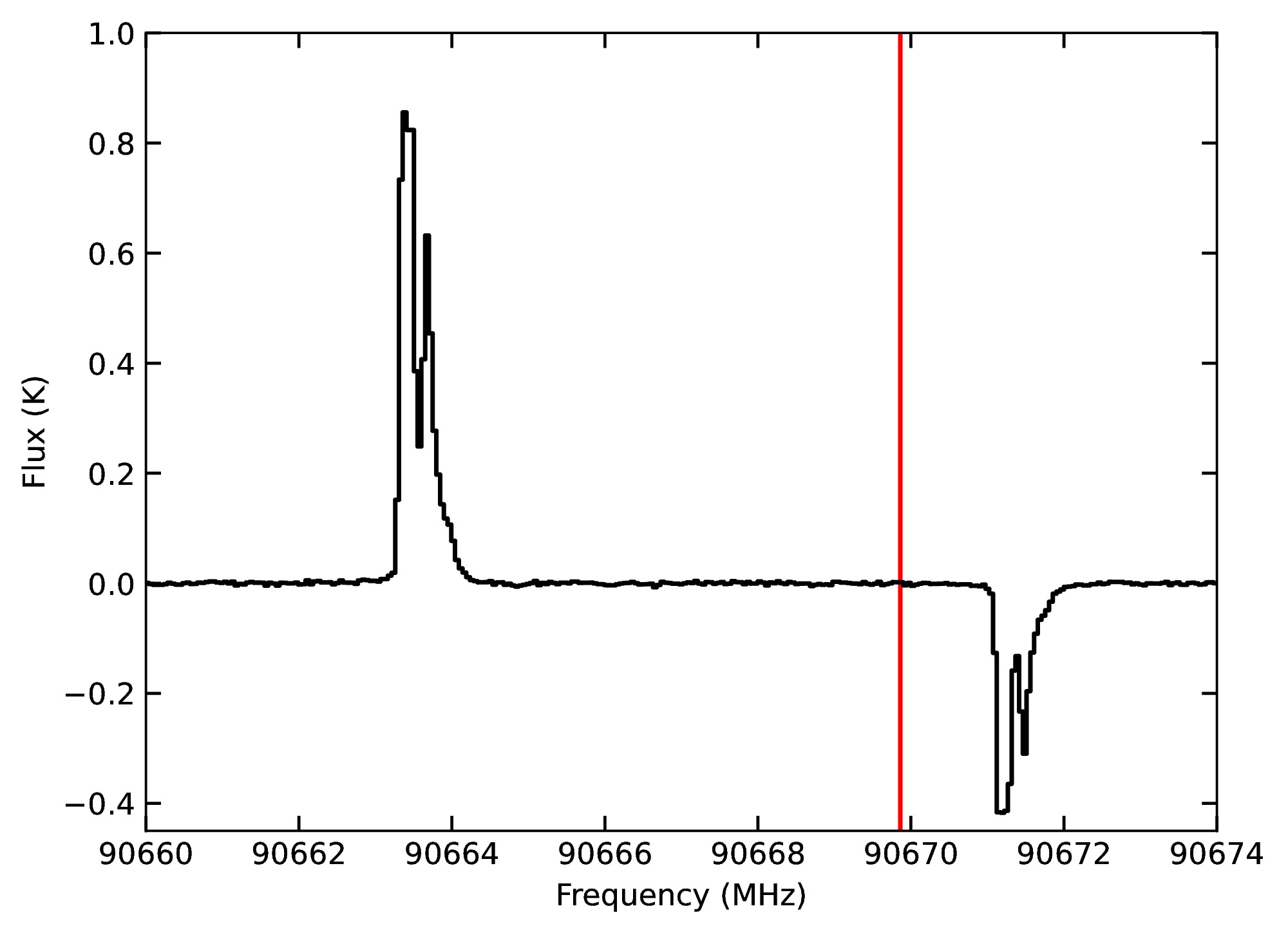}\includegraphics[width=0.49\textwidth]{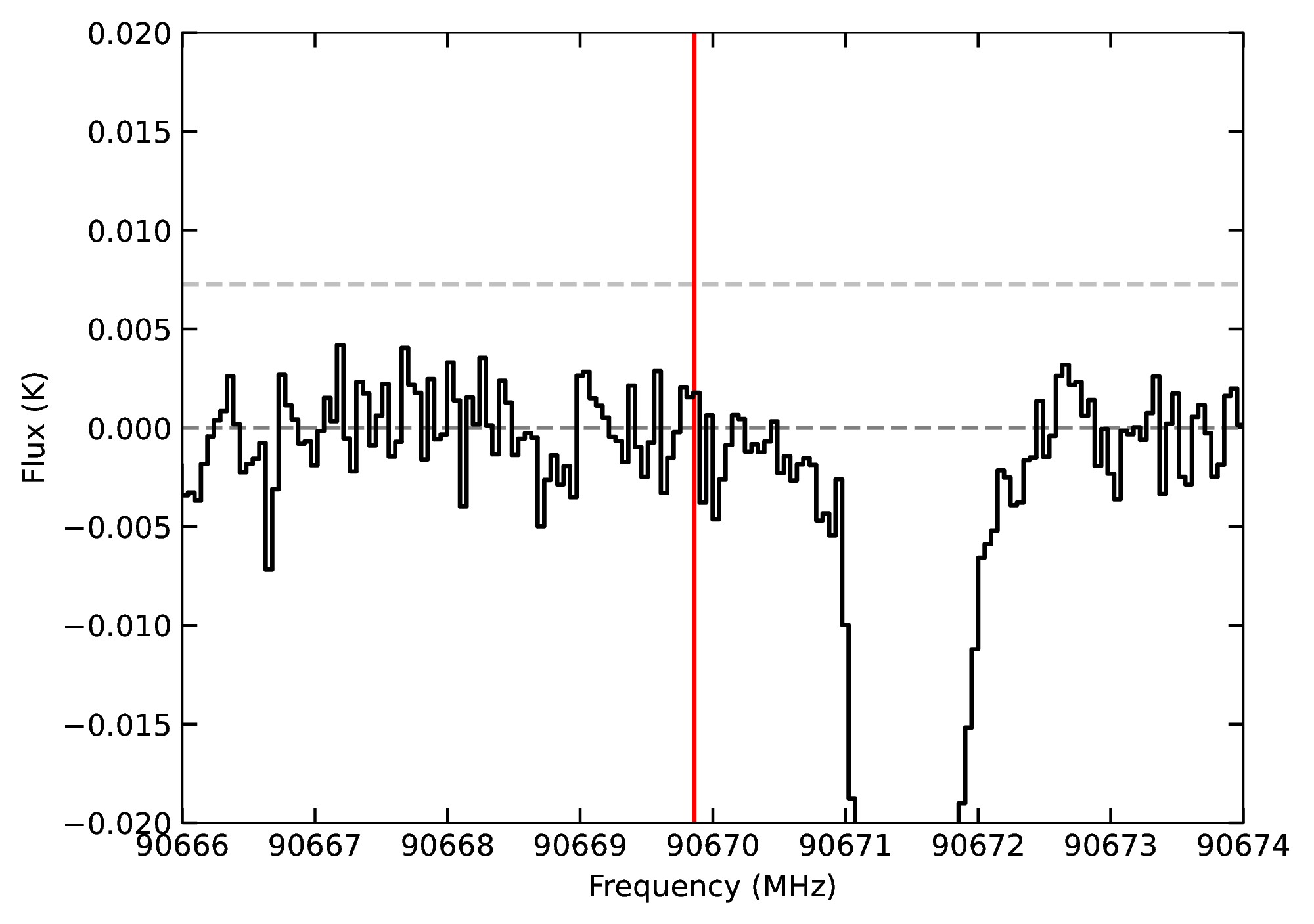}
    \caption{Same as in Fig. \ref{fig:freqthrow}, but for B213-C6.}
    \label{fig:freqthrow2}
\end{figure*}

\bibliography{References}{}

\begin{thebibliography}{}
\expandafter\ifx\csname natexlab\endcsname\relax\def\natexlab#1{#1}\fi
\providecommand{\url}[1]{\href{#1}{#1}}

\bibitem[{{Ag{\'u}ndez} {et~al.}(2019){Ag{\'u}ndez}, {Marcelino}, {Cernicharo}, {Roueff}, \& {Tafalla}}]{Agundez19}
{Ag{\'u}ndez}, M., {Marcelino}, N., {Cernicharo}, J., {Roueff}, E., \& {Tafalla}, M. 2019, \aap, 625, A147

\bibitem[{{Ambrose} {et~al.}(2021){Ambrose}, {Shirley}, \& {Scibelli}}]{Ambrose21}
{Ambrose}, H.~E., {Shirley}, Y.~L., \& {Scibelli}, S. 2021, \mnras, 501, 347

\bibitem[{{Andr{\'e}} {et~al.}(2014){Andr{\'e}}, {Di Francesco}, {Ward-Thompson}, {Inutsuka}, {Pudritz}, \& {Pineda}}]{Andre14}
{Andr{\'e}}, P., {Di Francesco}, J., {Ward-Thompson}, D., {et~al.} 2014, in Protostars and Planets VI, ed. H.~{Beuther}, R.~S. {Klessen}, C.~P. {Dullemond}, \& T.~{Henning}, 27--51

\bibitem[{{Bacmann} {et~al.}(2012){Bacmann}, {Taquet}, {Faure}, {Kahane}, \& {Ceccarelli}}]{Bacmann12}
{Bacmann}, A., {Taquet}, V., {Faure}, A., {Kahane}, C., \& {Ceccarelli}, C. 2012, \aap, 541, L12

\bibitem[{{Belloche} {et~al.}(2016){Belloche}, {M{\"u}ller}, {Garrod}, \& {Menten}}]{Belloche16}
{Belloche}, A., {M{\"u}ller}, H.~S.~P., {Garrod}, R.~T., \& {Menten}, K.~M. 2016, \aap, 587, A91

\bibitem[{{Benson} \& {Myers}(1989)}]{Benson89}
{Benson}, P.~J., \& {Myers}, P.~C. 1989, \apjs, 71, 89

\bibitem[{{Bergin} \& {Tafalla}(2007)}]{Bergin07}
{Bergin}, E.~A., \& {Tafalla}, M. 2007, \araa, 45, 339

\bibitem[{{Bergner} {et~al.}(2017){Bergner}, {{\"O}berg}, \& {Rajappan}}]{Bergner17}
{Bergner}, J.~B., {{\"O}berg}, K.~I., \& {Rajappan}, M. 2017, \apj, 845, 29

\bibitem[{{Bianchi} {et~al.}(2017{\natexlab{a}}){Bianchi}, {Codella}, {Ceccarelli}, {Taquet}, {Cabrit}, {Bacciotti}, {Bachiller}, {Chapillon}, {Gueth}, {Gusdorf}, {Lefloch}, {Leurini}, {Podio}, {Rygl}, {Tabone}, \& {Tafalla}}]{Bianchi17HH212}
{Bianchi}, E., {Codella}, C., {Ceccarelli}, C., {et~al.} 2017{\natexlab{a}}, \aap, 606, L7

\bibitem[{{Bianchi} {et~al.}(2017{\natexlab{b}}){Bianchi}, {Codella}, {Ceccarelli}, {Fontani}, {Testi}, {Bachiller}, {Lefloch}, {Podio}, \& {Taquet}}]{Bianchi17SVS13}
---. 2017{\natexlab{b}}, \mnras, 467, 3011

\bibitem[{{Bizzocchi} {et~al.}(2014){Bizzocchi}, {Caselli}, {Spezzano}, \& {Leonardo}}]{Bizzocchi14}
{Bizzocchi}, L., {Caselli}, P., {Spezzano}, S., \& {Leonardo}, E. 2014, \aap, 569, A27

\bibitem[{{B{\o}gelund} {et~al.}(2018){B{\o}gelund}, {McGuire}, {Ligterink}, {Taquet}, {Brogan}, {Hunter}, {Pearson}, {Hogerheijde}, \& {van Dishoeck}}]{Bögelund18}
{B{\o}gelund}, E.~G., {McGuire}, B.~A., {Ligterink}, N. F.~W., {et~al.} 2018, \aap, 615, A88

\bibitem[{{Boogert} {et~al.}(2015){Boogert}, {Gerakines}, \& {Whittet}}]{Boogert15}
{Boogert}, A.~C.~A., {Gerakines}, P.~A., \& {Whittet}, D. C.~B. 2015, \araa, 53, 541

\bibitem[{{Booth} {et~al.}(2021){Booth}, {Walsh}, {Terwisscha van Scheltinga}, {van Dishoeck}, {Ilee}, {Hogerheijde}, {Kama}, \& {Nomura}}]{Booth21}
{Booth}, A.~S., {Walsh}, C., {Terwisscha van Scheltinga}, J., {et~al.} 2021, Nature Astronomy, 5, 684

\bibitem[{{Caselli} \& {Ceccarelli}(2012)}]{Caselli12}
{Caselli}, P., \& {Ceccarelli}, C. 2012, \aapr, 20, 56

\bibitem[{{Chac{\'o}n-Tanarro} {et~al.}(2019){Chac{\'o}n-Tanarro}, {Caselli}, {Bizzocchi}, {Pineda}, {Sipil{\"a}}, {Vasyunin}, {Spezzano}, {Punanova}, {Giuliano}, \& {Lattanzi}}]{ChaconTanarro19}
{Chac{\'o}n-Tanarro}, A., {Caselli}, P., {Bizzocchi}, L., {et~al.} 2019, \aap, 622, A141

\bibitem[{{Chahine} {et~al.}(2022){Chahine}, {L{\'o}pez-Sepulcre}, {Neri}, {Ceccarelli}, {Mercimek}, {Codella}, {Bouvier}, {Bianchi}, {Favre}, {Podio}, {Alves}, {Sakai}, \& {Yamamoto}}]{Chahine22}
{Chahine}, L., {L{\'o}pez-Sepulcre}, A., {Neri}, R., {et~al.} 2022, \aap, 657, A78

\bibitem[{{Chuang} {et~al.}(2016){Chuang}, {Fedoseev}, {Ioppolo}, {van Dishoeck}, \& {Linnartz}}]{Chuang16}
{Chuang}, K.~J., {Fedoseev}, G., {Ioppolo}, S., {van Dishoeck}, E.~F., \& {Linnartz}, H. 2016, \mnras, 455, 1702

\bibitem[{{Codella} {et~al.}(2012){Codella}, {Ceccarelli}, {Lefloch}, {Fontani}, {Busquet}, {Caselli}, {Kahane}, {Lis}, {Taquet}, {Vasta}, {Viti}, \& {Wiesenfeld}}]{Codella12}
{Codella}, C., {Ceccarelli}, C., {Lefloch}, B., {et~al.} 2012, \apjl, 757, L9

\bibitem[{{Dartois} {et~al.}(2000){Dartois}, {Gerin}, \& {d'Hendecourt}}]{Dartois00}
{Dartois}, E., {Gerin}, M., \& {d'Hendecourt}, L. 2000, \aap, 361, 1095

\bibitem[{{Drozdovskaya} {et~al.}(2022){Drozdovskaya}, {Coudert}, {Margul{\`e}s}, {Coutens}, {J{\o}rgensen}, \& {Manigand}}]{Drozdovskaya22}
{Drozdovskaya}, M.~N., {Coudert}, L.~H., {Margul{\`e}s}, L., {et~al.} 2022, \aap, 659, A69

\bibitem[{{Drozdovskaya} {et~al.}(2021){Drozdovskaya}, {Schroeder I}, {Rubin}, {Altwegg}, {van Dishoeck}, {Kulterer}, {De Keyser}, {Fuselier}, \& {Combi}}]{Drozdovskaya21}
{Drozdovskaya}, M.~N., {Schroeder I}, I. R.~H.~G., {Rubin}, M., {et~al.} 2021, \mnras, 500, 4901

\bibitem[{{Endres} {et~al.}(2016){Endres}, {Schlemmer}, {Schilke}, {Stutzki}, \& {M{\"u}ller}}]{Endres16}
{Endres}, C.~P., {Schlemmer}, S., {Schilke}, P., {Stutzki}, J., \& {M{\"u}ller}, H. S.~P. 2016, Journal of Molecular Spectroscopy, 327, 95

\bibitem[{{Esplugues} {et~al.}(2022){Esplugues}, {Fuente}, {Navarro-Almaida}, {Rodr{\'\i}guez-Baras}, {Majumdar}, {Caselli}, {Wakelam}, {Roueff}, {Bachiller}, {Spezzano}, {Rivi{\`e}re-Marichalar}, {Mart{\'\i}n-Dom{\'e}nech}, \& {Mu{\~n}oz Caro}}]{Esplugues22}
{Esplugues}, G., {Fuente}, A., {Navarro-Almaida}, D., {et~al.} 2022, \aap, 662, A52

\bibitem[{{Fontani} {et~al.}(2015){Fontani}, {Busquet}, {Palau}, {Caselli}, {S{\'a}nchez-Monge}, {Tan}, \& {Audard}}]{Fontani15}
{Fontani}, F., {Busquet}, G., {Palau}, A., {et~al.} 2015, \aap, 575, A87

\bibitem[{{Fuchs} {et~al.}(2009){Fuchs}, {Cuppen}, {Ioppolo}, {Romanzin}, {Bisschop}, {Andersson}, {van Dishoeck}, \& {Linnartz}}]{Fuchs09}
{Fuchs}, G.~W., {Cuppen}, H.~M., {Ioppolo}, S., {et~al.} 2009, \aap, 505, 629

\bibitem[{{Fuente} {et~al.}(2019){Fuente}, {Navarro}, {Caselli}, {Gerin}, {Kramer}, {Roueff}, {Alonso-Albi}, {Bachiller}, {Cazaux}, {Commercon}, {Friesen}, {Garc{\'\i}a-Burillo}, {Giuliano}, {Goicoechea}, {Gratier}, {Hacar}, {Jim{\'e}nez-Serra}, {Kirk}, {Lattanzi}, {Loison}, {Malinen}, {Marcelino}, {Mart{\'\i}n-Dom{\'e}nech}, {Mu{\~n}oz-Caro}, {Pineda}, {Tafalla}, {Tercero}, {Ward-Thompson}, {Trevi{\~n}o-Morales}, {Rivi{\'e}re-Marichalar}, {Roncero}, {Vidal}, \& {Ballester}}]{Fuente19}
{Fuente}, A., {Navarro}, D.~G., {Caselli}, P., {et~al.} 2019, \aap, 624, A105

\bibitem[{{Geppert} {et~al.}(2006){Geppert}, {Hamberg}, {Thomas}, {{\"O}sterdahl}, {Hellberg}, {Zhaunerchyk}, {Ehlerding}, {Millar}, {Roberts}, {Semaniak}, {Ugglas}, {K{\"a}llberg}, {Simonsson}, {Kaminska}, \& {Larsson}}]{Geppert06}
{Geppert}, W.~D., {Hamberg}, M., {Thomas}, R.~D., {et~al.} 2006, Faraday Discussions, 133, 177

\bibitem[{{Herbst} \& {van Dishoeck}(2009)}]{Herbst09}
{Herbst}, E., \& {van Dishoeck}, E.~F. 2009, \araa, 47, 427

\bibitem[{{Hidaka} {et~al.}(2009){Hidaka}, {Watanabe}, {Kouchi}, \& {Watanabe}}]{Hidaka09}
{Hidaka}, H., {Watanabe}, M., {Kouchi}, A., \& {Watanabe}, N. 2009, \apj, 702, 291

\bibitem[{{Hsu} {et~al.}(2020){Hsu}, {Liu}, {Liu}, {Sahu}, {Hirano}, {Lee}, {Tatematsu}, {Kim}, {Juvela}, {Sanhueza}, {He}, {Johnstone}, {Qin}, {Bronfman}, {Chen}, {Dutta}, {Eden}, {Jhan}, {Kim}, {Kuan}, {Kwon}, {Lee}, {Lee}, {Moraghan}, {Rawlings}, {Shang}, {Soam}, {Thompson}, {Traficante}, {Wu}, {Yang}, \& {Zhang}}]{Hsu20}
{Hsu}, S.-Y., {Liu}, S.-Y., {Liu}, T., {et~al.} 2020, \apj, 898, 107

\bibitem[{{Hunter}(2007)}]{Hunter07}
{Hunter}, J.~D. 2007, Computing in Science and Engineering, 9, 90

\bibitem[{{Ilyushin} {et~al.}(2023){Ilyushin}, {M{\"u}ller}, {J{\o}rgensen}, {Bauerecker}, {Maul}, {Porohovoi}, {Alekseev}, {Dorovskaya}, {Lewen}, {Schlemmer}, \& {Lees}}]{Ilyushin23}
{Ilyushin}, V.~V., {M{\"u}ller}, H.~S.~P., {J{\o}rgensen}, J.~K., {et~al.} 2023, \aap, 677, A49

\bibitem[{{Ilyushin} {et~al.}(2024){Ilyushin}, {M{\"u}ller}, {Drozdovskaya}, {J{\o}rgensen}, {Bauerecker}, {Maul}, {Porohovoi}, {Alekseev}, {Dorovskaya}, {Zakharenko}, {Lewen}, {Schlemmer}, {Xu}, \& {Lees}}]{Ilyushin24}
{Ilyushin}, V.~V., {M{\"u}ller}, H.~S.~P., {Drozdovskaya}, M.~N., {et~al.} 2024, \aap, 687, A220

\bibitem[{{Imai} {et~al.}(2022){Imai}, {Oya}, {Svoboda}, {Liu}, {Lefloch}, {Viti}, {Zhang}, {Ceccarelli}, {Codella}, {Chandler}, {Sakai}, {Aikawa}, {Alves}, {Balucani}, {Bianchi}, {Bouvier}, {Busquet}, {Caselli}, {Caux}, {Charnley}, {Choudhury}, {Cuello}, {De Simone}, {Dulieu}, {Dur{\'a}n}, {Evans}, {Favre}, {Fedele}, {Feng}, {Fontani}, {Francis}, {Hama}, {Hanawa}, {Herbst}, {Hirano}, {Hirota}, {Isella}, {J{\'\i}menez-Serra}, {Johnstone}, {Kahane}, {Le Gal}, {Loinard}, {L{\'o}pez-Sepulcre}, {Maud}, {Maureira}, {Menard}, {Mercimek}, {Miotello}, {Moellenbrock}, {Mori}, {Murillo}, {Nakatani}, {Nomura}, {Oba}, {O'Donoghue}, {Ohashi}, {Okoda}, {Ospina-Zamudio}, {Pineda}, {Podio}, {Rimola}, {Sakai}, {Segura-Cox}, {Shirley}, {Taquet}, {Testi}, {Vastel}, {Watanabe}, {Watanabe}, {Witzel}, {Xue}, {Zhao}, \& {Yamamoto}}]{Imai22}
{Imai}, M., {Oya}, Y., {Svoboda}, B., {et~al.} 2022, \apj, 934, 70

\bibitem[{{Jim{\'e}nez-Serra} {et~al.}(2016){Jim{\'e}nez-Serra}, {Vasyunin}, {Caselli}, {Marcelino}, {Billot}, {Viti}, {Testi}, {Vastel}, {Lefloch}, \& {Bachiller}}]{JimenezSerra16}
{Jim{\'e}nez-Serra}, I., {Vasyunin}, A.~I., {Caselli}, P., {et~al.} 2016, \apjl, 830, L6

\bibitem[{{J{\o}rgensen} {et~al.}(2018){J{\o}rgensen}, {M{\"u}ller}, {Calcutt}, {Coutens}, {Drozdovskaya}, {{\"O}berg}, {Persson}, {Taquet}, {van Dishoeck}, \& {Wampfler}}]{Jorgensen18}
{J{\o}rgensen}, J.~K., {M{\"u}ller}, H.~S.~P., {Calcutt}, H., {et~al.} 2018, \aap, 620, A170

\bibitem[{{Kulterer} {et~al.}(2022){Kulterer}, {Drozdovskaya}, {Antonellini}, {Walsh}, \& {Millar}}]{Kulterer22}
{Kulterer}, B.~M., {Drozdovskaya}, M.~N., {Antonellini}, S., {Walsh}, C., \& {Millar}, T.~J. 2022, ACS Earth and Space Chemistry, 6, 1171

\bibitem[{{Lada} {et~al.}(1996){Lada}, {Alves}, \& {Lada}}]{Lada96}
{Lada}, C.~J., {Alves}, J., \& {Lada}, E.~A. 1996, \aj, 111, 1964

\bibitem[{{Lattanzi} {et~al.}(2020){Lattanzi}, {Bizzocchi}, {Vasyunin}, {Harju}, {Giuliano}, {Vastel}, \& {Caselli}}]{Lattanzi20}
{Lattanzi}, V., {Bizzocchi}, L., {Vasyunin}, A.~I., {et~al.} 2020, \aap, 633, A118

\bibitem[{{Lee} {et~al.}(2019){Lee}, {Codella}, {Li}, \& {Liu}}]{Lee19}
{Lee}, C.-F., {Codella}, C., {Li}, Z.-Y., \& {Liu}, S.-Y. 2019, \apj, 876, 63

\bibitem[{{Lee} {et~al.}(2023){Lee}, {Baek}, {Lee}, {Jeong}, {Kim}, {Aikawa}, {Herczeg}, {Johnstone}, \& {Tobin}}]{Lee23}
{Lee}, J.-E., {Baek}, G., {Lee}, S., {et~al.} 2023, \apj, 956, 43

\bibitem[{{Ligterink} {et~al.}(2021){Ligterink}, {Ahmadi}, {Coutens}, {Tychoniec}, {Calcutt}, {van Dishoeck}, {Linnartz}, {J{\o}rgensen}, {Garrod}, \& {Bouwman}}]{Ligterink21}
{Ligterink}, N.~F.~W., {Ahmadi}, A., {Coutens}, A., {et~al.} 2021, \aap, 647, A87

\bibitem[{{Lin} {et~al.}(2023){Lin}, {Spezzano}, \& {Caselli}}]{Lin23}
{Lin}, Y., {Spezzano}, S., \& {Caselli}, P. 2023, \aap, 669, L6

\bibitem[{{Linsky} {et~al.}(2006){Linsky}, {Draine}, {Moos}, {Jenkins}, {Wood}, {Oliveira}, {Blair}, {Friedman}, {Gry}, {Knauth}, {Kruk}, {Lacour}, {Lehner}, {Redfield}, {Shull}, {Sonneborn}, \& {Williger}}]{Linsky06}
{Linsky}, J.~L., {Draine}, B.~T., {Moos}, H.~W., {et~al.} 2006, \apj, 647, 1106

\bibitem[{{Luhman} {et~al.}(2009){Luhman}, {Mamajek}, {Allen}, \& {Cruz}}]{Luhman09}
{Luhman}, K.~L., {Mamajek}, E.~E., {Allen}, P.~R., \& {Cruz}, K.~L. 2009, \apj, 703, 399

\bibitem[{{Luhman} {et~al.}(2003){Luhman}, {Stauffer}, {Muench}, {Rieke}, {Lada}, {Bouvier}, \& {Lada}}]{Luhman03}
{Luhman}, K.~L., {Stauffer}, J.~R., {Muench}, A.~A., {et~al.} 2003, \apj, 593, 1093

\bibitem[{{Manigand} {et~al.}(2020){Manigand}, {J{\o}rgensen}, {Calcutt}, {M{\"u}ller}, {Ligterink}, {Coutens}, {Drozdovskaya}, {van Dishoeck}, \& {Wampfler}}]{Manigand20}
{Manigand}, S., {J{\o}rgensen}, J.~K., {Calcutt}, H., {et~al.} 2020, \aap, 635, A48

\bibitem[{{Mart{\'\i}n-Dom{\'e}nech} {et~al.}(2021){Mart{\'\i}n-Dom{\'e}nech}, {Bergner}, {{\"O}berg}, {Carpenter}, {Law}, {Huang}, {J{\o}rgensen}, {Schwarz}, \& {Wilner}}]{MartinDomenech21}
{Mart{\'\i}n-Dom{\'e}nech}, R., {Bergner}, J.~B., {{\"O}berg}, K.~I., {et~al.} 2021, \apj, 923, 155

\bibitem[{{Meg{\'\i}as} {et~al.}(2023){Meg{\'\i}as}, {Jim{\'e}nez-Serra}, {Mart{\'\i}n-Pintado}, {Vasyunin}, {Spezzano}, {Caselli}, {Cosentino}, \& {Viti}}]{Megias23}
{Meg{\'\i}as}, A., {Jim{\'e}nez-Serra}, I., {Mart{\'\i}n-Pintado}, J., {et~al.} 2023, \mnras, 519, 1601

\bibitem[{{Mercimek} {et~al.}(2022){Mercimek}, {Codella}, {Podio}, {Bianchi}, {Chahine}, {Bouvier}, {L{\'o}pez-Sepulcre}, {Neri}, \& {Ceccarelli}}]{Mercimek22}
{Mercimek}, S., {Codella}, C., {Podio}, L., {et~al.} 2022, \aap, 659, A67

\bibitem[{{Millar} {et~al.}(1989){Millar}, {Bennett}, \& {Herbst}}]{Millar89}
{Millar}, T.~J., {Bennett}, A., \& {Herbst}, E. 1989, \apj, 340, 906

\bibitem[{{M{\"u}ller} {et~al.}(2005){M{\"u}ller}, {Schl{\"o}der}, {Stutzki}, \& {Winnewisser}}]{Muller05}
{M{\"u}ller}, H. S.~P., {Schl{\"o}der}, F., {Stutzki}, J., \& {Winnewisser}, G. 2005, Journal of Molecular Structure, 742, 215

\bibitem[{{M{\"u}ller} {et~al.}(2001){M{\"u}ller}, {Thorwirth}, {Roth}, \& {Winnewisser}}]{Muller01}
{M{\"u}ller}, H.~S.~P., {Thorwirth}, S., {Roth}, D.~A., \& {Winnewisser}, G. 2001, \aap, 370, L49

\bibitem[{{Nagaoka} {et~al.}(2005){Nagaoka}, {Watanabe}, \& {Kouchi}}]{Nagaoka05}
{Nagaoka}, A., {Watanabe}, N., \& {Kouchi}, A. 2005, \apjl, 624, L29

\bibitem[{{Neill} {et~al.}(2013){Neill}, {Crockett}, {Bergin}, {Pearson}, \& {Xu}}]{Neill13}
{Neill}, J.~L., {Crockett}, N.~R., {Bergin}, E.~A., {Pearson}, J.~C., \& {Xu}, L.-H. 2013, \apj, 777, 85

\bibitem[{{Nomura} {et~al.}(2022){Nomura}, {Furuya}, {Cordiner}, {Charnley}, {Alexander}, {Nixon}, {Guzman}, {Yurimoto}, {Tsukagoshi}, \& {Iino}}]{Nomura22}
{Nomura}, H., {Furuya}, K., {Cordiner}, M.~A., {et~al.} 2022, arXiv e-prints, arXiv:2203.10863

\bibitem[{{{\"O}berg} {et~al.}(2009){{\"O}berg}, {Garrod}, {van Dishoeck}, \& {Linnartz}}]{Oberg09}
{{\"O}berg}, K.~I., {Garrod}, R.~T., {van Dishoeck}, E.~F., \& {Linnartz}, H. 2009, \aap, 504, 891

\bibitem[{{Okoda} {et~al.}(2024){Okoda}, {Oya}, {Sakai}, {Watanabe}, {L{\'o}pez-Sepulcre}, {Oyama}, {Zeng}, \& {Yamamoto}}]{Okoda24}
{Okoda}, Y., {Oya}, Y., {Sakai}, N., {et~al.} 2024, \apj, 970, 28

\bibitem[{{Okoda} {et~al.}(2023){Okoda}, {Oya}, {Francis}, {Johnstone}, {Ceccarelli}, {Codella}, {Chandler}, {Sakai}, {Aikawa}, {Alves}, {Herbst}, {Maureira}, {Bouvier}, {Caselli}, {Choudhury}, {De Simone}, {J{\'\i}menez-Serra}, {Pineda}, \& {Yamamoto}}]{Okoda23}
{Okoda}, Y., {Oya}, Y., {Francis}, L., {et~al.} 2023, \apj, 948, 127

\bibitem[{{Osamura} {et~al.}(2004){Osamura}, {Roberts}, \& {Herbst}}]{Osamura04}
{Osamura}, Y., {Roberts}, H., \& {Herbst}, E. 2004, \aap, 421, 1101

\bibitem[{{Ospina-Zamudio} {et~al.}(2018){Ospina-Zamudio}, {Lefloch}, {Ceccarelli}, {Kahane}, {Favre}, {L{\'o}pez-Sepulcre}, \& {Montarges}}]{OspinaZamudio18}
{Ospina-Zamudio}, J., {Lefloch}, B., {Ceccarelli}, C., {et~al.} 2018, \aap, 618, A145

\bibitem[{{Ospina-Zamudio} {et~al.}(2019){Ospina-Zamudio}, {Favre}, {Kounkel}, {Xu}, {Neill}, {Lefloch}, {Faure}, {Bergin}, {Fedele}, \& {Hartmann}}]{OspinaZamudio19}
{Ospina-Zamudio}, J., {Favre}, C., {Kounkel}, M., {et~al.} 2019, \aap, 627, A80

\bibitem[{{Parise} {et~al.}(2004){Parise}, {Castets}, {Herbst}, {Caux}, {Ceccarelli}, {Mukhopadhyay}, \& {Tielens}}]{Parise04}
{Parise}, B., {Castets}, A., {Herbst}, E., {et~al.} 2004, \aap, 416, 159

\bibitem[{{Parise} {et~al.}(2006){Parise}, {Ceccarelli}, {Tielens}, {Castets}, {Caux}, {Lefloch}, \& {Maret}}]{Parise06}
{Parise}, B., {Ceccarelli}, C., {Tielens}, A.~G.~G.~M., {et~al.} 2006, \aap, 453, 949

\bibitem[{{Parise} {et~al.}(2002){Parise}, {Ceccarelli}, {Tielens}, {Herbst}, {Lefloch}, {Caux}, {Castets}, {Mukhopadhyay}, {Pagani}, \& {Loinard}}]{Parise02}
---. 2002, \aap, 393, L49

\bibitem[{Pearson {et~al.}(2012)Pearson, Yu, \& Drouin}]{PEARSON2012119}
Pearson, J.~C., Yu, S., \& Drouin, B.~J. 2012, Journal of Molecular Spectroscopy, 280, 119, broadband Rotational Spectroscopy.
\newblock \url{https://www.sciencedirect.com/science/article/pii/S0022285212001026}

\bibitem[{{Peng} {et~al.}(2012){Peng}, {Despois}, {Brouillet}, {Parise}, \& {Baudry}}]{Peng12}
{Peng}, T.~C., {Despois}, D., {Brouillet}, N., {Parise}, B., \& {Baudry}, A. 2012, \aap, 543, A152

\bibitem[{{Pickett} {et~al.}(1998){Pickett}, {Poynter}, {Cohen}, {Delitsky}, {Pearson}, \& {M{\"u}ller}}]{Pickett98}
{Pickett}, H.~M., {Poynter}, R.~L., {Cohen}, E.~A., {et~al.} 1998, \jqsrt, 60, 883

\bibitem[{{Qasim} {et~al.}(2018){Qasim}, {Chuang}, {Fedoseev}, {Ioppolo}, {Boogert}, \& {Linnartz}}]{Qasim18}
{Qasim}, D., {Chuang}, K.~J., {Fedoseev}, G., {et~al.} 2018, \aap, 612, A83

\bibitem[{{Ratajczak} {et~al.}(2009){Ratajczak}, {Quirico}, {Faure}, {Schmitt}, \& {Ceccarelli}}]{Ratacjzak09}
{Ratajczak}, A., {Quirico}, E., {Faure}, A., {Schmitt}, B., \& {Ceccarelli}, C. 2009, \aap, 496, L21

\bibitem[{{Rebull} {et~al.}(2010){Rebull}, {Padgett}, {McCabe}, {Hillenbrand}, {Stapelfeldt}, {Noriega-Crespo}, {Carey}, {Brooke}, {Huard}, {Terebey}, {Audard}, {Monin}, {Fukagawa}, {G{\"u}del}, {Knapp}, {Menard}, {Allen}, {Angione}, {Baldovin-Saavedra}, {Bouvier}, {Briggs}, {Dougados}, {Evans}, {Flagey}, {Guieu}, {Grosso}, {Glauser}, {Harvey}, {Hines}, {Latter}, {Skinner}, {Strom}, {Tromp}, \& {Wolf}}]{Rebull2010}
{Rebull}, L.~M., {Padgett}, D.~L., {McCabe}, C.~E., {et~al.} 2010, \apjs, 186, 259

\bibitem[{{Riedel} {et~al.}(2023){Riedel}, {Sipil{\"a}}, {Redaelli}, {Caselli}, {Vasyunin}, {Dulieu}, \& {Watanabe}}]{Riedel23}
{Riedel}, W., {Sipil{\"a}}, O., {Redaelli}, E., {et~al.} 2023, \aap, 680, A87

\bibitem[{{Rodr{\'\i}guez-Baras} {et~al.}(2021){Rodr{\'\i}guez-Baras}, {Fuente}, {Rivi{\'e}re-Marichalar}, {Navarro-Almaida}, {Caselli}, {Gerin}, {Kramer}, {Roueff}, {Wakelam}, {Esplugues}, {Garc{\'\i}a-Burillo}, {Le Gal}, {Spezzano}, {Alonso-Albi}, {Bachiller}, {Cazaux}, {Commercon}, {Goicoechea}, {Loison}, {Trevi{\~n}o-Morales}, {Roncero}, {Jim{\'e}nez-Serra}, {Laas}, {Hacar}, {Kirk}, {Lattanzi}, {Mart{\'\i}n-Dom{\'e}nech}, {Mu{\~n}oz-Caro}, {Pineda}, {Tercero}, {Ward-Thompson}, {Tafalla}, {Marcelino}, {Malinen}, {Friesen}, \& {Giuliano}}]{RodriguezBaras21}
{Rodr{\'\i}guez-Baras}, M., {Fuente}, A., {Rivi{\'e}re-Marichalar}, P., {et~al.} 2021, \aap, 648, A120

\bibitem[{{Rodr{\'\i}guez-Baras} {et~al.}(2023){Rodr{\'\i}guez-Baras}, {Esplugues}, {Fuente}, {Spezzano}, {Caselli}, {Loison}, {Roueff}, {Navarro-Almaida}, {Bachiller}, {Mart{\'\i}n-Dom{\'e}nech}, {Jim{\'e}nez-Serra}, {Beitia-Antero}, \& {Le Gal}}]{Rodriguez-Baras23}
{Rodr{\'\i}guez-Baras}, M., {Esplugues}, G., {Fuente}, A., {et~al.} 2023, \aap, 679, A120

\bibitem[{{Sahu} {et~al.}(2019){Sahu}, {Liu}, {Su}, {Li}, {Lee}, {Hirano}, \& {Takakuwa}}]{Sahu19}
{Sahu}, D., {Liu}, S.-Y., {Su}, Y.-N., {et~al.} 2019, \apj, 872, 196

\bibitem[{{Santos} {et~al.}(2022){Santos}, {Chuang}, {Lamberts}, {Fedoseev}, {Ioppolo}, \& {Linnartz}}]{Santos22}
{Santos}, J.~C., {Chuang}, K.-J., {Lamberts}, T., {et~al.} 2022, \apjl, 931, L33

\bibitem[{{Scibelli} {et~al.}(2025){Scibelli}, {Drozdovskaya}, {Caselli}, {Ferrer Asensio}, {Kulterer}, {Spezzano}, {Lin}, \& {Shirley}}]{Scibelli25}
{Scibelli}, S., {Drozdovskaya}, M.~N., {Caselli}, P., {et~al.} 2025, arXiv e-prints, arXiv:2508.04762

\bibitem[{{Scir{\`e}} {et~al.}(2019){Scir{\`e}}, {Urso}, {Fulvio}, {Baratta}, \& {Palumbo}}]{Scire19}
{Scir{\`e}}, C., {Urso}, R.~G., {Fulvio}, D., {Baratta}, G.~A., \& {Palumbo}, M.~E. 2019, Spectrochimica Acta Part A: Molecular Spectroscopy, 219, 288

\bibitem[{{Slavicinska} {et~al.}(2024){Slavicinska}, {van Dishoeck}, {Tychoniec}, {Nazari}, {Rubinstein}, {Gutermuth}, {Tyagi}, {Chen}, {Brunken}, {Rocha}, {Manoj}, {Narang}, {Thomas Megeath}, {Yang}, {Looney}, {Tobin}, {Beuther}, {Bourke}, {Linnartz}, {Federman}, {Watson}, \& {Linz}}]{Slavicinska24}
{Slavicinska}, K., {van Dishoeck}, E.~F., {Tychoniec}, {\L}., {et~al.} 2024, \aap, 688, A29

\bibitem[{{Spezzano} {et~al.}(2022){Spezzano}, {Fuente}, {Caselli}, {Vasyunin}, {Navarro-Almaida}, {Rodr{\'\i}guez-Baras}, {Punanova}, {Vastel}, \& {Wakelam}}]{Spezzano22}
{Spezzano}, S., {Fuente}, A., {Caselli}, P., {et~al.} 2022, \aap, 657, A10

\bibitem[{{Taquet} {et~al.}(2012){Taquet}, {Ceccarelli}, \& {Kahane}}]{Taquet12}
{Taquet}, V., {Ceccarelli}, C., \& {Kahane}, C. 2012, \apjl, 748, L3

\bibitem[{{Taquet} {et~al.}(2019){Taquet}, {Bianchi}, {Codella}, {Persson}, {Ceccarelli}, {Cabrit}, {J{\o}rgensen}, {Kahane}, {L{\'o}pez-Sepulcre}, \& {Neri}}]{Taquet19}
{Taquet}, V., {Bianchi}, E., {Codella}, C., {et~al.} 2019, \aap, 632, A19

\bibitem[{{van der Walt} {et~al.}(2011){van der Walt}, {Colbert}, \& {Varoquaux}}]{vanderWalt11}
{van der Walt}, S., {Colbert}, S.~C., \& {Varoquaux}, G. 2011, Computing in Science and Engineering, 13, 22

\bibitem[{{van der Walt} {et~al.}(2023){van der Walt}, {Kristensen}, {Calcutt}, {J{\o}rgensen}, \& {Garrod}}]{vanderWalt23}
{van der Walt}, S.~J., {Kristensen}, L.~E., {Calcutt}, H., {J{\o}rgensen}, J.~K., \& {Garrod}, R.~T. 2023, \aap, 677, A127

\bibitem[{{van Gelder} {et~al.}(2020){van Gelder}, {Tabone}, {Tychoniec}, {van Dishoeck}, {Beuther}, {Boogert}, {Caratti o Garatti}, {Klaassen}, {Linnartz}, {M{\"u}ller}, \& {Taquet}}]{vanGelder20}
{van Gelder}, M.~L., {Tabone}, B., {Tychoniec}, {\L}., {et~al.} 2020, \aap, 639, A87

\bibitem[{{van Gelder} {et~al.}(2022){van Gelder}, {Jaspers}, {Nazari}, {Ahmadi}, {van Dishoeck}, {Beltr{\'a}n}, {Fuller}, {S{\'a}nchez-Monge}, \& {Schilke}}]{vanGelder22}
{van Gelder}, M.~L., {Jaspers}, J., {Nazari}, P., {et~al.} 2022, \aap, 667, A136

\bibitem[{{Vastel} {et~al.}(2015){Vastel}, {Bottinelli}, {Caux}, {Glorian}, \& {Boiziot}}]{Vastel15}
{Vastel}, C., {Bottinelli}, S., {Caux}, E., {Glorian}, J.~M., \& {Boiziot}, M. 2015, in SF2A-2015: Proceedings of the Annual meeting of the French Society of Astronomy and Astrophysics, 313--316

\bibitem[{Virtanen {et~al.}(2020)Virtanen, Gommers, Oliphant, Haberland, Reddy, Cournapeau, Burovski, Peterson, Weckesser, Bright, {van der Walt}, Brett, Wilson, Millman, Mayorov, Nelson, Jones, Kern, Larson, Carey, Polat, Feng, Moore, {VanderPlas}, Laxalde, Perktold, Cimrman, Henriksen, Quintero, Harris, Archibald, Ribeiro, Pedregosa, {van Mulbregt}, \& {SciPy 1.0 Contributors}}]{scipy}
Virtanen, P., Gommers, R., Oliphant, T.~E., {et~al.} 2020, Nature Methods, 17, 261

\bibitem[{{Watanabe} \& {Kouchi}(2002)}]{Watanabe02}
{Watanabe}, N., \& {Kouchi}, A. 2002, \apjl, 571, L173

\bibitem[{{Watson}(1974)}]{Watson74}
{Watson}, W.~D. 1974, \apj, 188, 35

\bibitem[{{Wilkins} \& {Blake}(2022)}]{Wilkins22}
{Wilkins}, O.~H., \& {Blake}, G.~A. 2022, Journal of Physical Chemistry A, 126, 6473

\end{thebibliography}
\bibliographystyle{aasjournal}

\end{document}